\documentclass[12pt]{article}
\pdfoutput=1

\usepackage{draft} 
\usepackage{hyperref}
\usepackage{graphicx,color}
\usepackage{cite}
\usepackage{mciteplus}
\usepackage{skak}
\begin{document}

\unitlength = .8mm

\begin{titlepage}
\rightline{MIT-CTP-4648}

\begin{center}

\hfill \\
\hfill \\
\vskip 1cm

\title{On Higher Derivative Couplings\\  in Theories with Sixteen Supersymmetries}

\author{Ying-Hsuan Lin$^\textsymrook$, Shu-Heng Shao$^\textsymrook$, Yifan Wang$^\textsymknight$, Xi Yin$^\textsymrook$}

\address{
$^\textsymrook$Jefferson Physical Laboratory, Harvard University, \\
Cambridge, MA 02138 USA
\\
$^\textsymknight$Center for Theoretical Physics, Massachusetts Institute of Technology, \\
Cambridge, MA 02139 USA}

\email{yhlin@physics.harvard.edu, shuhengshao@fas.harvard.edu, yifanw@mit.edu,
xiyin@fas.harvard.edu}

\end{center}

\abstract{ We give simple arguments for new non-renormalization theorems on higher derivative couplings of gauge theories to supergravity, with sixteen supersymmetries, by considerations of brane-bulk superamplitudes. This leads to some exact results on the effective coupling of D3-branes in type IIB string theory. We also derive exact results on   higher dimensional operators in the torus compactification of the six dimensional $(0,2)$ superconformal theory. }

\vfill

\end{titlepage}

\eject

\tableofcontents

\section{Introduction}

A great deal of the dynamics of maximally supersymmetric gauge theories and string theories can be learned from the derivative expansion of the effective action, in appropriate phases where the low energy description is simple. On the other hand, it is often nontrivial to implement the full constraints of supersymmetry on the dynamics, due to the lack of a convenient superspace formalism that makes 16 or 32 supersymmetries manifest (see \cite{Howe:1980th,Howe:1981xy,Berkovits:1997pj,Cederwall:2001dx,Bossard:2010bd,Chang:2014kma,Chang:2014nwa} however for on-shell superspace and pure spinor superspace approaches). It became clear recently \cite{Brodel:2009hu,Elvang:2010jv,Elvang:2010xn,Wang:2015jna} that on-shell supervertices and scattering amplitudes can be used to organize higher derivative couplings efficiently in maximally supersymmetric theories, and highly nontrivial renormalization theorems of \cite{Sethi:1999qv,Green:1998by} can be argued in a remarkably simple way based on considerations of amplitudes. 

In this paper we extend the arguments of \cite{Wang:2015jna} to gauge theories coupled to maximal supergravity, while preserving 16 supersymmetries. Our primary example is an Abelian gauge theory on a 3-brane coupled to ten dimensional type IIB supergravity, though the strategy may be applied to other dimensions as well. We will formulate in detail the brane-bulk superamplitudes, utilizing the super spinor helicity formalism in four dimensions \cite{Elvang:2013cua} as well as in  type IIB supergravity \cite{CaronHuot:2010rj,Boels:2012ie}. By considerations of local supervertices, and factorization of nonlocal superamplitudes, we will derive constraints on the higher derivative brane-bulk couplings of the form $F^4$, $RF^2, D^2 RF^2, D^4RF^2, R^2, D^2 R^2$.  These amount to a set of non-renormalization theorems, which when combined with $SL(2,\mathbb{Z})$ invariance, determines the $\tau,\bar\tau$ dependence of such couplings completely in the quantum effective action of a D3 brane in type IIB string theory. Some of these results have previously been observed through explicit string theory computations \cite{Bachas:1999um,Green:2000ke,Fotopoulos:2001pt,Fotopoulos:2002wy,Basu:2008gt,Garousi:2010ki,Garousi:2010bm,Garousi:2011ut,Becker:2011ar,Becker:2011bw,Velni:2012sv,Garousi:2012yr}.

We then turn to the question of determining higher dimensional operators that appear in the four dimensional gauge theory obtained by compactifying the six dimensional $(0,2)$ superconformal theory on a torus. While it is unclear whether this theory can be coupled to the ten dimensional type IIB supergravity, we will be able to derive nontrivial constraints and an exact result on the $F^4$ term by interpolating the effective theory in the Coulomb phase, and matching with perturbative double scaled little string theory. Our result clarifies some puzzles that previously existed in the literature.

\section{Brane-Bulk Superamplitudes}

We begin by considering a maximally supersymmetric Abelian gauge multiplet on a 3-brane coupled to type IIB supergravity in ten dimensions. 
The super spinor helicity variables of the ten dimensional type IIB supergravity multiplet are $\zeta_{\underline{\A} A}$ and $\eta_A$, where $\underline{\A}=1,\cdots,16$ is an $SO(1,9)$ chiral spinor index, and $A=1,\cdots 8$ is an $SO(8)$ little group chiral spinor index. The spinor helicity variables $\zeta_{\underline{\A} A}$ are constrained via the null momentum $p_m$ by 
\ie
\delta_{AB} p_m = \Gamma_m^{\underline{\A}\underline{\B}} \zeta_{\underline{\A} A} \zeta_{\underline{\B} B}.
\fe
A 1-particle state in the type IIB supergravity multiplet is labeled by a monomial in $\eta_A$. For instance, 1 and $\eta^8\equiv {1\over 8!}\epsilon_{A_1\cdots A_8}\eta_{A_1}\cdots \eta_{A_8}$ correspond to the axion-dilaton fields $\tau$ and $\bar\tau$, $\eta_{[A}\eta_{B]}$ and ${1\over 6!}\epsilon_{AB A_1\cdots A_6} \eta_{A_1}\cdots \eta_{A_6}$ correspond to the complexified 2-form fields, and $\eta_{[A}\eta_B \eta_C \eta_{D]}$ contains the graviton and the self-dual 4-form.
The 32 supercharges ${\bf q}_{\underline{\A}}, {\bf \widetilde q}_{\underline{\A}}$ act on the 1-particle states as \cite{Boels:2012ie}
\ie
{\bf q}_{\underline{\A}} = \zeta_{\underline{\A} A} \eta_A,~~~~ {\bf \overline q}_{\underline{\A}} = \zeta_{\underline{\A} A} {\partial\over \partial \eta_A}.
\fe
The supersymmetry algebra takes the form
\ie
\{ {\bf q}_{\underline{\A}}, {\bf q}_{\underline{\B}} \} = \{  {\bf \overline q}_{\underline{\A}} ,  {\bf \overline q}_{\underline{\B}}  \} = 0,~~~~ \{{\bf q}_{\underline{\A}}, {\bf \overline q}_{\underline{\B}}  \} = {1\over 2} p_m \Gamma^m_{\underline{\A}\underline{\B}}.
\fe
To describe coupling to the brane, let us decompose the supercharges with respect to $SO(1,3)\times SO(6)$, and write
\ie
{\bf q}_{\underline\A} = (q_{\A I}, \widetilde q_{\da}{}^I),~~~~ {\bf \overline q}_{\underline\A} = (\overline{\widetilde q}_{\A I}, \overline{ q}{}_{\da}{}^I).
\fe
Here $\A$ and $\da$ are four dimensional chiral and anti-chiral spinor indices, and the lower and upper index $I$ label the chiral and anti-chiral spinors of $SO(6)$. The coupling to four dimensional gauge multiplet on the brane will preserve 16 out of the 32 supercharges, which we take to be $q_{aI}$ and $\overline{ q}{}_{\da}{}^I$.

%One may be tempted to take the preserved supercharges to be $q_{\A I}$ and $\overline{\widetilde q}{}_{\da}{}^I$, and as in the usual formulation of four dimensional superamplitudes, one would treat $q_{\A I}$ as supermomenta and write $\overline{\widetilde q}{}_{\da}{}^I$ as superderivatives on a set of Grassmannian variables $\theta_I$. This formalism has the advantage of making the $SO(6)$ R-symmetry in the four dimensional theory manifest, but is inconvenient for writing superamplitudes that exhibit the correct little group scaling.

The four dimensional super spinor helicity variables for the gauge multiplet are $\lambda_\A, \widetilde\lambda_\db, \theta_I$. The null momentum and supercharges of a particle in the multiplet are given by \cite{Elvang:2013cua}
\ie
p_\mu = \sigma_\mu^{\A\db} \lambda_\A \widetilde \lambda_{\db} ,
~~~
q_{\A I} = \lambda_\A \theta_I, ~~~ \overline q_{\db}{}^I = \widetilde \lambda_{\db} {\partial\over \partial \theta_I}.
\fe
The $SO(2)$ little group acts by
\ie\label{sot}
\lambda\to e^{i\A} \lambda, ~~~\widetilde\lambda\to e^{-i\A}\widetilde\lambda, ~~~\theta\to e^{-i\A} \theta.
\fe
Here we adopt a slightly unconventional little group transformation of $\theta_I$, so that $q_{\A I}, \widetilde q_{\db}{}^I$ are invariant under the little group, and can be combined with the supermomenta of the bulk supergravitons in constructing a superamplitude. A 1-particle state in a gauge multiplet is represented by a monomial in $\theta_I$. For instance, 1 and $\theta^4\equiv {1\over 4!}\epsilon^{IJKL}\theta_I\theta_J\theta_K\theta_L$ represent the $-$ and $+$ helicity gauge bosons,\footnote{Note that our sign convention for helicity is the opposite of \cite{Elvang:2013cua}.} while $\theta_I \theta_J$ represent the scalar field $\phi_{[IJ]}$.

%In order to couple to the 3+1 dimensional gauge multiplet, we identify $(q_{aI}, \widetilde q_{\dot b}{}^I)$ with half of the 32 ten dimensional supercharges via the decomposition
%\ie
%& {\bf q}_\A = \big( q_{aI}, {q^\ddagger}_{\dot b}{}^I \big),
%~~~
%{\bf \widetilde q}_\A = \big( {\widetilde q^\ddagger}{}_{aI}, \widetilde q_{\dot b}{}^I \big).
%\fe
%Our convention on the little group transformation of $\theta_I$ allows for combining the supermomenta of the fields on the brane with those of the fields in the bulk, and will be convenient for constructing amplitudes that obey supersymmetry Ward identities.

In an $n$-point superamplitude that involves particles in the four dimensional gauge multiplet as well as the ten dimensional gravity multiplet, only the four dimensional momentum $P_\mu = \sum_{i=1}^n p_{i\mu}$ and the 16 supercharges $(Q_{\A I},\overline Q_{\db}{}^I) $ are conserved. Here we have defined
\ie
&Q_{\A I} =\sum_{i=1}^n q_{i\A I}= \sum_i \lambda_{i\A} \theta_{iI} + \sum_j \xi_{j\A I A} \eta_{jA},\\
&\overline Q_{\db}{}^I =\sum_{i=1}^n\overline q_{i\db}{}^I= \sum_i \widetilde\lambda_{i\db} {\partial\over \partial \theta_{iI}} + \sum_j \widetilde\xi_{j\db}{}^I{}_A {\partial\over \partial \eta_{jA}},
\fe
where $\widetilde \xi_{i\db}{}^I$ is
the decomposition of the supergravity spinor helicity variable $\zeta_{i\underline{\A} A}$ with respect to $SO(1,3)\times SO(6)\subset SO(1,9)$, namely
% the decomposition of $\zeta_{i\A A}$ and $\eta_{iA}$ with respect to $SO(3,1)\times SU(4)\subset SO(9,1)\times SO(8)$, that is,
\ie
%& \zeta_{i,\A, A} = (\zeta'_{i,aI,J}, \zeta_{i,aI,}{}^J, \widetilde \zeta_{i,\dot b}{}^{I,}{}_{J},\widetilde \zeta'_{i,\dot b}{}^{I,J}),~~~~ \eta_{iA} = (\eta_{iI}, \widetilde \eta_i{}^I)
& \zeta_{i\underline{\A} A} = (\xi_{i\A I A},\widetilde \xi_{i\db}{}^I{}_A).
\fe

 A typical superamplitude takes the form\footnote{The only exceptions are when the kinematics are constrained in such a way that no nontrivial Lorentz and little group invariants can be formed, such as the 3-graviton amplitude in the bulk, the graviton tadpole on the brane, and the graviton-gauge multiplet coupling on the brane. These will be examined in more detail below.}
\ie\label{ampa}
{\cal A} = \delta^4(P_\mu) \delta^8(Q_{\A I}) {\cal F}(\lambda_i,\widetilde\lambda_i,\theta_i, \zeta_j,\eta_j),
\fe
where
\ie
\D^8(Q_{\A I})\equiv \prod_{\A,I} Q_{\A I},
\fe
and ${\cal F}$  obeys supersymmetry Ward identities \cite{Elvang:2010xn}
\ie
\delta^4(P_\mu) \delta^8(Q_{\A I}) \,\overline Q_{\db}{}^J {\cal F} = 0
\fe
associated with the 8 $\overline Q$ supercharges. 

If the amplitude $\mathcal{A}$ (\ref{ampa}) obeys supersymmetry Ward identities, then so does its CPT conjugate
\ie
\overline{\cal A} = \delta^4(P_\mu) \overline Q^8 {\cal F}(\lambda_i,\widetilde\lambda_i,{\partial/\partial\theta_i}, \zeta_j,{\partial/\partial\eta_j}) \prod_i \theta_i^4 \prod_j \eta_j^8,
\fe
where $\overline Q^8\equiv \prod_{\da, I} \overline Q_\da{}^I$.

In formulating superamplitudes purely in the gauge theory, it is useful to work with a different representation of the 16 supercharges, by decomposing
\ie\label{splitpreservedQ}
Q_{\A I} =(\mathcal{Q}_{\A a}, \mathcal{ \overline Q}_{\A \dot a}),~~~~{\overline{ Q}}_{\da}{}^I = (\mathcal{\overline{ Q}}{}_{\da a}, { \mathcal{Q}}{}_{\da\dot a}),
\fe
where $(a,\dot a)$ are spinor indices of an $SU(2)\times SU(2)$ subgroup of the $SU(4)$ R-symmetry. We can then represent the supercharges for individual particles through Grassmannian variables $(\psi_a, \widetilde\psi_{\dot a})$ as
%\ie
%& \mathfrak{q}_{\A a} = \lambda_\A \psi_a,~~~\overline{\mathfrak{q}}_{\A\dot a} = \lambda_\A {\partial\over \partial \widetilde\psi^{\dot a}},
%\\
%& \mathfrak{\overline{ q}}{}_{\da a} = \widetilde\lambda_\da {\partial\over \partial\psi^a},~~~ \mathfrak{q}_{\da\dot a} = \widetilde \lambda_\da \widetilde\psi_{\dot a},
%\fe
%where ${\cal Q}=\sum_i \mathfrak q_i$ and ${\cal \overline Q}=\sum_i \overline{\mathfrak q}_i$ as before.
\ie
& \mathcal{Q}_{\A a} = \lambda_\A \psi_a,~~~\mathcal{\overline Q}_{\A\dot a} = \lambda_\A {\partial\over \partial \widetilde\psi^{\dot a}},
\\
& \mathcal{\overline{ Q}}{}_{\da a} = \widetilde\lambda_\da {\partial\over \partial\psi^a},~~~ \mathcal{Q}_{\da\dot a} = \widetilde \lambda_\da \widetilde\psi_{\dot a}.
\fe

In this representation, a basis of 1-particle states is given by monomials in $\psi,\widetilde\psi$. The $-$ and $+$ helicity gauge bosons correspond to $\psi^2$ and $\widetilde\psi^2$, whereas the scalars are represented by 1, $\psi^2\widetilde\psi^2$, and $\psi_a\widetilde\psi_{\dot a}$. We can assign $\psi_a$ and $\widetilde\psi_{\dot a}$ to transform under the $SO(2)$ little group with charge $-1$ and $+1$, respectively.

The $\theta$-representation of superamplitude is convenient for coupling to supergravity, while the $\psi$-representation is convenient for constructing vertices of the gauge theory that solve supersymmetry Ward identities. The superamplitudes in the $\theta$-representation and in the $\psi$-representation are related by a Grassmannian twistor transform:
\ie
{\cal A}_\theta = \int \prod_i d^2\widetilde\psi_i \, e^{\sum_i \widetilde\psi_i \chi_i} {\cal A}_\psi,
\fe
where we make the identification $\theta_\A = (\psi_a, \chi_{\dot a})$, after picking an $SU(2)\times SU(2)$ subgroup of $SU(4)$ R-symmetry.

A typical supervertex constructed in the $\psi$-representation is not manifestly R-symmetry invariant. In a supervertex that involves bulk supergravitons, we can form R-symmetry invariant supervertices by contracting with the spinor helicity variable of the supergraviton, or simply its transverse momentum to the 3-brane, and average over the $SO(6)$ orbit. It is useful to record the non-manifest R-symmetry generators in the $\psi$-representation,
\ie
R_{a\dot b} = \sum_i \psi_{ia} \widetilde\psi_{i \dot b},~~~ R_{\dot a b} = \sum_i {\partial\over \partial \psi_i^a} {\partial\over \partial \widetilde\psi_i^{\dot b}},~~~ R = \sum_i \left(\psi_{ia} {\partial\over \partial \psi_{ia}} + \widetilde\psi_{i\dot a} {\partial\over \partial \widetilde\psi_{i\dot a}} - 2\right).
\fe

\subsection{F-term and D-term Supervertices}

Let us focus on supervertices, namely, local superamplitudes with no poles in momenta. As in maximal supergravity theories, we can write down F-term and D-term supervertices \cite{Wang:2015jna} for brane-bulk coupling. One may attempt to write construct a simple class of supervertices in the form (\ref{ampa}) by taking ${\cal F}$ to be independent of the Grassmann variables $\theta_i, \eta_j$, and depend only on the bosonic spinor helicity variables, subject to $SO(1,3)\times SO(6)$ invariance. When combined with the CPT conjugate vertex, this construction appears to be sufficiently general for purely gravitational F-term vertices. For instance, a supervertex involve 2 bulk supergravitons of the form
\ie
\delta^4(P)\delta^8(Q) = \delta^4(p_1^\parallel+p_2^\parallel) \delta^8(q_1+q_2)
\fe
corresponds to a coupling of the form $R^2+\cdots$ on the brane.

When there are four dimensional gauge multiplet particles involved, however, such simple constructions in the $\theta$-representation of the superamplitude may not give the correct little group scaling. It is sometimes more convenient to start with a supervertex in the $\psi$-representation, average over $SO(6)$, and perform the twistor transform into $\theta$-representation.
For instance, we can write a supervertex that involves $(4+n)$ gauge multiplet particles in the $\psi$-representation, of the form 
\ie
\delta^4(P) \delta^8(\mathcal{Q}_\psi)=\delta^4(P) \delta^8(\mathcal{Q}_{\A a}, \mathcal{Q}_{\da \dot a}) = \delta^4(P) \delta^4\left(\sum_{i=1}^{n+4} \lambda_{i\A} \psi_{ia} \right) \delta^4\left(\sum_{i=1}^{n+4} \widetilde\lambda_{i\da} \widetilde\psi_{i\dot a} \right).
\fe
This vertex is not $SO(6)$ invariant; rather, it lies in the lowest weight component of a rank $n$ symmetric traceless tensor representation of the $SO(6)$ R-symmetry. In component fields, it contains couplings of the form $\phi^{i_1}\cdots \phi^{i_{n}} F^4+\cdots$, where $\phi^i$ denotes the 6 scalars, and the traces between $i_k, i_\ell$ are subtracted off. 

Indeed, one can verify that for the 4-point superamplitude,
\ie
\int \prod_{i=1}^4 d^2\widetilde\psi_i\, e^{\sum_i \widetilde\psi_i \chi_i} \delta^4(P)\delta^8(\mathcal{Q}_\psi) = \delta^4(P) \delta^8(Q_\theta) {[34]^2\over \langle 12\rangle^2},
\fe
while the analogous twistor transform on $\delta^4(P)\delta^8(\mathcal{Q}_\psi)$ for $n>0$ produces $\delta^4(P)\delta^8(Q_\theta)$ multiplied by an expression of degree $2n$ in $\chi$, that transforms nontrivially under the $SO(6)$. It is generally more difficult to extend a gauge supervertex constructed in the $\psi$-representation to involve coupling to the supergraviton however. 

As an example, we construct supervertices in the $\psi$-representation which contain $\phi^m\partial^m R^2$ couplings on the brane. These supervertices are naturally related to the $R^2$ vertex by spontaneously broken translation symmetry. To proceed, we first need to extend the $\psi$-representation to the supergraviton states.

Just as we split the 16 preserved supercharges on the brane in \eqref{splitpreservedQ}, we can split the 16 broken supercharges as follows,
\ie
 \overline{\widetilde Q}_{\A I} = (\overline{\widetilde\cQ}_{\A a}, {\widetilde\cQ}_{\A \dot a}),~~~ \widetilde Q_\da{}^I = ({\widetilde\cQ} _{\da a}, \overline{\widetilde\cQ}_{\da\dot a}),
\fe
We'd like to consider a representation of the supergraviton states such that $(\cQ_{\A a}, \cQ_{\da\dot a}, \widetilde\cQ_{\A\dot a}, {\widetilde\cQ}_{\da a})$ are represented as supermomenta, and the remaining 16 supercharges are represented as superderivatives. This is possible provided that $(\cQ_{\A a}, \cQ_{\da\dot a}, \widetilde\cQ_{\A\dot a}, {\widetilde\cQ}_{\da a})$ anticommute with one another. The anticommutator of $Q_{\A I}$ with $\overline{\widetilde Q}_{\B J}$ contains the transverse momentum $P_{IJ}$. Hence while ${\cQ}_{\A a}$ anticommute with $\widetilde\cQ_{\db b}$, it may not anticommute with $\widetilde\cQ_{\B\dot b}$. However, the anticommutator $\{\cQ_{\A a}, \widetilde\cQ_{\B\dot b}\}$ contains only the component $P_{a\dot b}$ that lies in the representation $(2,2)^0$ through the decomposition $6\to (2,2)^0\oplus (1,1)^+\oplus (1,1)^-$ under $SU(2)\times SU(2)\times U(1)\subset SO(6)$. As long as there are no more than two supergravitons in the supervertex, we can always choose the $SO(4)$ subgroup of $SO(6)$ to leave the two transverse momenta of the supergravitons invariant so that $P_{a\dot b}=0$. With this choice, for each supergraviton, $(\cQ_{\A a}, \cQ_{\da\dot a}, \widetilde\cQ_{\A\dot a}, {\widetilde\cQ}_{\da a})$ then anti-commute with one another, and they can be simultaneously represented as supermomenta. 

Let us compare this with the standard representation of the supercharges in the 10D type IIB super spinor helicity formalism, for which we can decompose
\ie
\zeta_{\underline{\A} A} = (\zeta_{\A I A}; \zeta_{\da}{}^I{}_A) = (\zeta_{\A a A}, \zeta_{\A \dot a A}; \zeta_{\da a A}, \zeta_{\da\dot a A}).
\fe
%We have the identification
%\ie
%\cQ_{\A a} = \zeta_{\A a A}\eta_A,~~~ \overline{\cQ}_{\A\dot a} = \zeta_{\A\dot a A}\eta_A,~~~
%\widetilde\cQ_{\da a} = \zeta_{\da a A}\eta_A,~~~ \overline{\widetilde\cQ}_{\da \dot a} = \zeta_{\da \dot a A}\eta_A.
%\fe
%Now the supermomenta in the $\psi$-representation are supposed to be identified with
%\ie
%\cQ_{\A a} = \zeta_{\A a A}\eta_A,~~~ {\cQ}_{\da\dot a} = \zeta_{\da\dot a A}{\partial\over \partial\eta_A},~~~
%\widetilde\cQ_{\da a} = \zeta_{\da a A}\eta_A,~~~ {\widetilde\cQ}_{\A \dot a} = \zeta_{\A \dot a A}{\partial\over\partial\eta_A}.
%\fe
%This is only possible if the transverse momenta of the supergravitons are invariant under an $SO(4)$ subgroup of the $SO(6)$ rotation, as discussed above. In this case, 
By requiring that $P_{a\dot b}=0$, we have
\ie
\epsilon^{\A\B}\zeta_{\A a A}\zeta_{\B \dot b A} = \epsilon^{\da\db} \zeta_{\da a A} \zeta_{\db \dot b A}= 0.
\fe
When this condition is satisfied, we can go to the $\psi$-representation by a Laplace transform on half of the 8 $\eta_A$'s. 

A supervertex of the form
\ie
\delta^8(\cQ_{\A a}, \cQ_{\da\dot a}, \widetilde\cQ_{\A\dot a}, \widetilde{\cQ}_{\da a})
\fe
for 2 supergravitons and $m$ D3-brane gauge multiplets is not $SO(6)$ invariant (unless $m=0$). Rather, it lies in the lowest weight component in a set of supervertices that transform in the rank $m$ symmetric traceless representation of $SO(6)$. To form an $SO(6)$ invariant supervertex, we need to contract it with $m$ powers of the total transverse momentum $P_{IJ}$, and average over the $SO(6)$ orbit. In this way, we obtain the desire supervertex that contains $\phi^m \partial^m R^2$ coupling.

%In below, we consider $\theta$-representation ands write $Q$ for $Q_\theta$ when there is no room for confusion.
%
%
%There is a 3-point supervertex that couples one supergraviton to two gauge multiplet particles, that scales like the (transverse) momentum squared, which we may write in the $\theta$-representation, of the form
%\ie
%\delta^4(P) \delta^8(Q) {1\over \langle 12\rangle^2},
%\fe
%where $p_1, p_2$ are the momenta of the two gauge multiplet particles.

\subsection{Elementary Vertices}

\begin{figure}[htb]
\centering
\begin{minipage}{0.32\textwidth}
\centering
\includegraphics[scale=1.7]{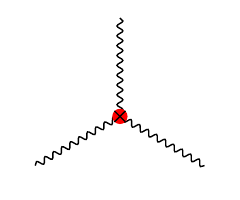}\\
$\mathcal{A}_3$
\end{minipage}  
\begin{minipage}{0.32\textwidth}
\centering
\includegraphics[scale=1.7 ]{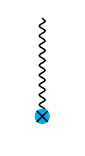}\\
$\mathcal{B}_1$
\end{minipage} 
\raisebox{-20pt}{
\begin{minipage}{0.32\textwidth}
\centering
\includegraphics[scale=1.7 ]{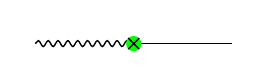}\\
$\mathcal{B}_{1,1}$
\end{minipage} 
}
\caption{Elementary supervertices. The wiggly line represents a bulk 1-particle state while the straight line represents a brane 1-particle state. The red dot represents the bulk vertex, whereas the blue and green dots are brane vertices.}
\label{elementary}
\end{figure}

There are a few ``elementary vertices" that are the basic building blocks of the brane coupling to supergravity, and are not of the form of the F and D-term vertices discussed above. One elementary vertex is the supergravity 3-point vertex (Figure \ref{elementary}), as discussed in \cite{Boels:2012ie}. In the notation of \cite{Wang:2015jna}, it can be written in the form
\ie
{\cal A}_3={g\over (p_+)^4}\delta^{10}(P)\delta^{12}(W),
\fe
where $g$ is the cubic coupling constant, $W$ represents 12 independent components of the supermomentum, specified by the null plane that contains the three external null momenta, and $p^+$ is an overall lightcone momentum as defined in \cite{Wang:2015jna}. The explicit expression of this vertex will not be discussed here, though the cubic vertex is of course crucial in the consideration of factorization of superamplitudes.

The supergraviton tadpole on the brane is a 1-point superamplitude, of the form
\ie
{\cal B}_1 = T \delta^4(P) \Pi^{ABCD}(\zeta)\eta_A\eta_B\eta_C\eta_D,
\fe
where $T$ stands for the tension/charge of the brane, and $\Pi^{ABCD}(\zeta)$ is an anti-symmetric 4-tensor of the $SO(8)$ little group constructed out of the $\zeta_{\A A}$ associated with a (complex) null momentum in the 6-plane transverse to the 3-brane, of homogeneous degree zero in $\zeta$. If we take the transverse momentum to be in a lightcone direction, after double Wick rotation, the little group $SO(8)$ transverse to the lightcone is broken by the 3-brane to $SO(4)\times SO(4)$. We may then decompose $\eta_A = (\eta_{\A a}^+, \eta^-_{\da\dot a})$, where $(\A,\da)$ are spinor indices of the $SO(4)$ along the brane worldvolume, whereas $(a,\dot a)$ are spinor indices of the $SO(4)$ transverse to the brane as well as the null momentum. With respect to the $SO(4)\times SO(4)$, the 16 supercharges $Q_{\A I},\overline Q_{\db}{}^I$ preserved by the 3-brane coupling may be denoted $Q_{\A a}, Q_{\A\dot a}, \overline Q_{\db b}, \overline Q_{\db \dot b}$. $Q_{\A \dot a}$ and $\overline Q_{\db\dot b}$ trivially annihilate the 1-particle state of the supergraviton, $Q_{\A a}\sim \eta^+_{\A a}$, and $\overline Q_{\db\dot b} \sim \partial/\partial \eta^{-\db\dot b}$. The supergraviton tadpole supervertex can then be written as
\ie\label{tps}
{\cal B}_1 = T \delta^4(P) (\eta^+)^4.
\fe
This amplitude contains equal amount of graviton tadpole and the charge with respect to the 4-form potential, reflecting the familiar BPS relation between the tension and charge of the brane.

The supergraviton-gauge multiplet 2-point vertex $\mathcal{B}_{1,1}$ is another elementary vertex. Here again there is no Lorentz invariant to be formed out of the two external null momenta. Both the transverse and parallel components of the graviton momentum are null. To write this vertex explicitly, we take the graviton transverse momentum to be along a lightlike direction on the $(X^8,X^9)$ plane, and the parallel momentum to be along a lightlike direction on the $(X^0,X^1)$ plane. We will write the  null parallel and transverse momenta $p^\parallel,\, p^\perp$ in this frame as
\ie
 &p^\parallel = ( p^{\parallel}_{+}  , \,p^{\parallel}_{ +},\,0 ,\cdots,0, 0),\\
 &p^\perp = (0 ,0,0\cdots, \,   i p^{\perp}_{+}  ,\, p^{\perp}_{+} ).
 \fe
 Note that $p^{\parallel}_{+}  ,\, p^{\perp}_{+} $ transform under the boosts on the $(X^0,X^1)$ and $(X^8,X^9)$ planes, which will be important for us to fix the $p^{\parallel}_{+}  ,\, p^{\perp}_{+} $ dependence in the supervertex $\mathcal{B}_{1,1}$.

 The ``tiny group" $SO(6)$ that acts on the transverse directions to the null plane spanned by the momenta of the two particles (one on the brane, one in the bulk) rotates $X^2,\cdots,X^7$, which is broken by the 3-brane to $SO(2)\times SO(4)$. The spinor helicity variables are decomposed as
\ie\label{frame}
& \xi_{\A I A} = \Big( \xi_{+a |A} ,\,\xi_{-a |A} ,\,\xi_{+\dot a |A} ,\,\xi_{-\dot a |A}=0 \Big),
\\
& \widetilde\xi_{\dot\A ~A}^{~I} = \Big(  \widetilde\xi_{+a |A} ,\, \widetilde\xi_{-a |A} ,\, \widetilde\xi_{+\dot a |A} ,\, \widetilde\xi_{-\dot a |A}=0 \Big),\\
& \lambda_\A = \Big(\lambda_+=\sqrt{p^{\parallel}_{+} },\lambda_-=0\Big),~~~\widetilde\lambda_\da = \Big(\widetilde\lambda_+=\sqrt{p^{\parallel}_{+} },\widetilde\lambda_- = 0\Big).
\fe
%\ie
%& \zeta_{\underline{\A} A} = \Big( \zeta_{+a |A} ,\,\zeta_{-a |A} ,\,\zeta_{+\dot a |A} ,\,\zeta_{-\dot a |A}=0 \Big),
%\\
%& \lambda_\A = \Big(\lambda_+=\sqrt{p_+},\lambda_-=0\Big),~~~\widetilde\lambda_\da = \Big(\widetilde\lambda_+=\sqrt{p_+},\widetilde\lambda_- = 0\Big).
%\fe
We will also split $\theta_I=(\theta_a, \theta_{\dot a})$.
The 16 unbroken supercharges are represented as 
\ie
& Q_{+a} = \xi_{+a|A} \eta_A + \lambda_+\theta_a, ~~ Q_{-a} = \xi_{-a|A}\eta_A ,~~Q_{+\dot a} = \xi_{+\dot a|A}\eta_A + \lambda_+ \theta_{\dot a},~~Q_{-\dot a} = 0,
\\
& \overline Q_{+,a} =\widetilde \xi_{+a|A} {\partial\over \partial\eta_A} + \widetilde\lambda_+ {\partial\over \partial \theta^a},~~ \overline Q_{-,a} = \widetilde\xi_{-a|A} {\partial\over \partial\eta_A},
~~ \overline Q_{+,\dot a} = \widetilde\xi_{+\dot a|A} {\partial\over \partial\eta_A} + \widetilde\lambda_+ {\partial\over \partial \theta^{\dot a}},~~\overline Q_{-,\dot a} = 0 .
\fe
The supervertex can be written in this frame as\footnote{This supervertex is very similar to the cubic vertex in the non-Abelian gauge theory, which is absent here because we  restrict to the Abelian case. 
}
\ie
{\cal B}_{1,1} = \sqrt{Tg} \delta^4(P) {\delta^{6}(Q_{+a}, Q_{-a}, Q_{+\dot a})\over p^{\parallel}_{+}  p^{\perp}_{+} },
\fe
From  boost invariance on the $(X^0,X^1)$ plane, we know there is one power of $p^{\parallel}_{+} $  in the denominator. Since the supervertex scales linearly with the momentum, we determine the factor $p^{\perp}_{+} $ in the denominator.

\begin{figure}[htb]
\centering
\begin{minipage}{0.49\textwidth}
\centering
\includegraphics[scale=1.8 ]{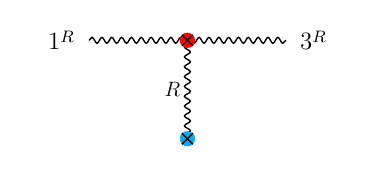}
\end{minipage}  
\begin{minipage}{0.49\textwidth}
\centering
\includegraphics[scale=1.8 ]{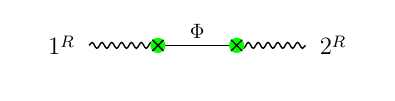}
\end{minipage} 
\caption{Factorization of the $R^2$ amplitude through elementary vertices. The red dot represents the bulk supergravity vertex whereas the blue and green dots are brane vertices.}
\label{RR}
\end{figure}

 The normalization of $\mathcal{B}_{1,1}$ is unambiguously fixed by supersymmetry. Note that there is a unique 2-supergraviton amplitude of the form \cite{Hashimoto:1996bf}
\ie\label{qst}
{\delta^8(Q)\over st},
\fe
 at this order in  momentum. Here $s=-(p_1+p_2)^2$, $t=(p_1^\perp)^2=(p_2^\perp)^2$. The 2-supergraviton amplitude  factorizes through ${\cal B}_1 {\cal A}_3$ and ${\cal B}_{1,1}{\cal B}_{1,1}$ (Figure \ref{RR}), from which the relative coefficients of these two channels are fixed (proportional to $Tg$). %Note that in (\ref{qst}) the residue in $s$ and in $t$ cannot be adjusted independently, while maintaining the same momentum scaling and simple pole condition.

\subsection{Examples of Superamplitudes}

\begin{figure}[htb]
\centering
{
\begin{minipage}{0.32\textwidth}
\centering
\includegraphics[scale=2 ]{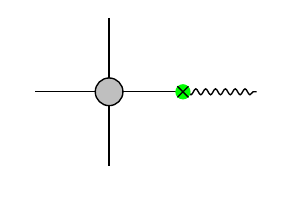}\\
\end{minipage} 
}
\caption{A factorization for the $RF^3$ superamplitude for the case of an Abelian gauge multiplet coupled to supergravity.}
\label{fig:RF3}
\end{figure}

Let us now attempt to construct a 4-point superamplitude that couples one supergraviton to three gauge multiplet particles, that scales like $p^3$ (Figure~\ref{fig:RF3}). We will see that such a superamplitude must be nonlocal, and an independent local supervertex of this form does not exist. This superamplitude should be of the form $\delta^4(P) \delta^8(Q)$ times a rational function that has total degree 2 in $\eta$ and $\theta$,\footnote{If the vertex, at the same derivative order, is $\delta^4(P)\delta^8(Q)$ times a function of $\lambda,\widetilde\lambda$ that is independent of $\eta$ and $\theta$, this function needs to have homogeneous degree $-4$ in the $\lambda$'s and degree 2 in the $\widetilde\lambda$'s in order to reproduce the correct little group scaling. Such a function cannot be a polynomial in the spinor helicity variables and the amplitude would have to be nonlocal. The situation is similar if the function is of degree 4 in $\eta$ and $\theta$, which may be obtained from the CPT conjugate of the previous case. It seems that such amplitudes cannot factorize correctly into lower point supervertices (and they do not exist in string theory).} homogeneous degree $-1$ in the momenta, and must have the little group scaling such that a term $\sim\eta^4\theta_1^2\theta_2^2\theta_3^2$ (representing three scalars coupled to the graviton or the 4-form potential) is little group invariant.

To construct this superamplitude, we will pick the supergraviton momentum to be in the $X^9$ direction, and decompose the spinor helicity variables according to $SO(3)\times SO(5)\subset SO(8)$, where the $SO(8)$ that rotates $X^1,\cdots,X^8$ can be identified with the little group of the supergraviton, and the $SO(3)$ and $SO(5)$ rotate $X^1,X^2,X^3$ along the 3-brane and $X^4,\cdots,X^8$ transverse to the 3-brane, respectively. We can write 
$\eta_A = \eta_{\A I}$, where $\A$ is an $SO(3)$ spinor index and $I$ an $SO(5)$ spinor index. We can split $\zeta_{\underline{\A} A}$ into $(\zeta_{B A}, \zeta_{\dot B A})$, where $B$ and $\dot B$ are chiral and anti-chiral $SO(8)$ indices. Then the spinor helicity constraint on $\zeta$ is simply that $\zeta_{\dot B A} =0$, and $\zeta_{B A} = \sqrt{p^\perp} \delta_{B A}$. Further decomposing the index $B$ into $SO(3)\times SO(5)$ indices $\B J$, and identifying $A\sim \A I$, we have
\ie
\zeta_{\B J,\A I} = \sqrt{p^\perp} \epsilon_{\B\A} \Omega_{JI},
\fe
where $\Omega_{IJ}$ is the invariant anti-symmetric tensor of $SO(5)\sim Sp(4)$. The supercharges can now be written explicitly (in $SO(3)\times SO(5)$ notation) as
\ie\label{supq}
& Q_{\A I} = \sqrt{p^\perp} \eta_{\A I} + \sum_{i=1}^3 \lambda_{i\A} \theta_{iI},
\\
& \overline Q_{\A I} = \sqrt{p^\perp} {\partial\over \partial \eta^{\A I}} + \sum_{i=1}^3 \widetilde\lambda_{i\A} {\partial\over \partial \theta_i^I}.
\fe
The general superamplitude that solves the supersymmetry Ward identity and has the correct little group scaling and momentum takes the form
\ie\label{paq}
\delta^4(P) \delta^8(Q) \sum_{i,j}% {\langle ij\rangle r_{ij}(s,t,u)\over \langle 23\rangle \langle 34\rangle \langle 42\rangle }
f_{ij}(\lambda_k,\widetilde\lambda_k) \left(\widetilde\lambda_{i\A}\eta^{\A}{}_I - \sqrt{p^\perp} \theta_{iI}\right) \left(\widetilde\lambda_{j\B}\eta^\B{}_J - \sqrt{p^\perp} \theta_{jJ}\right) \Omega^{IJ},
\fe
where $f_{ij}$ is a rational function of $\lambda_{k\A}$ and $\widetilde\lambda_{k\A}$, $k=1,2,3$. Note that since we are working in a frame tied to the supergraviton momentum, $\A$ is an $SO(3)$ index, and we can contract $\lambda_i$ with $\widetilde\lambda_j$, and write for instance $[j i\rangle = \widetilde\lambda_{j\A}\lambda_i^\A$. The little group and momentum scaling demands that $f_{ij}$ has homogeneous degree $-4$ in the $\lambda_k$'s and degree 0 in the $\widetilde\lambda_k$'s.

Due to the $\delta^8(Q)$ factor, we can rewrite (\ref{paq}) as
\ie\label{pqn}
\delta^4(P) \delta^8(Q) (p^\perp)^{-1} \sum_{i,j}
f_{ij}(\lambda,\widetilde\lambda) \big([ik\rangle \theta_{kI} + p^\perp \theta_{iI}\big) \big([j\ell \rangle \theta_{\ell J} + p^\perp \theta_{jJ}\big) \Omega^{IJ}.
\fe
It appears that such an amplitude with the correct little group scaling will necessarily have poles, thereby forbidding a local supervertex.\footnote{Note that we can shift
\ie
f_{ij}(\lambda,\widetilde\lambda) \to f_{ij}(\lambda,\widetilde\lambda) + \lambda_{i\A} g^\A_j + \lambda_{j\A} g_i^\A
\fe
for arbitrary $g_i^\A$ without changing the amplitude (\ref{pqn}). }

The corresponding 4-point disc amplitude on D3-brane in type IIB string theory has a pole in $(p^\perp)^2$, and no pole in $s,t,u$ (at zero value). Here $s=-(p_1+p_2)^2$, $t=-(p_2+p_3)^2$, $u=-(p_3+p_1)^2$, with $s+t+u=(p^\perp)^2$. In particular, there is a coupling $(\partial_i \delta\overline\tau) \phi^i F_-^2 $, that corresponds to the term proportional to $\eta^8\theta_{iI}\theta_{iJ}\Omega^{IJ}$ in (\ref{pqn}). This coupling is represented by
\ie
\eta^8 p^\perp \left( [12]^2\theta_{3I}\theta_{3J} + [23]^2\theta_{1I}\theta_{1J} + [31]^2\theta_{2I}\theta_{2J} \right) \Omega^{IJ}
\fe
Comparing to (\ref{pqn}), we need
\ie
\sum_{i,j} f_{ij} [i1\rangle [j 1\rangle + 2 p^\perp \sum_i f_{1i}  [i1\rangle +  (p^\perp)^2 f_{11} = {[23]^2\over (p^\perp)^2}.
\fe
A solution for $f_{ij}$ with the correct little group scaling is
\ie
& f_{11} =  {[23]^2 \over (p^\perp)^4}, ~~~ f_{22} = {[31]^2 \over (p^\perp)^4},~~~f_{33} =  {[12]^2 \over (p^\perp)^4},
\\
& f_{12} = -{[13][23] \over (p^\perp)^4}, ~~~ f_{23} = -{[21][31] \over (p^\perp)^4},~~~f_{31} = -{[32][12] \over (p^\perp)^4}.
\fe
%\ie
%& f_{11} =- {[23]^2 \over (p^\perp)^4}, ~~~ f_{22} =- {[31]^2 \over (p^\perp)^4},~~~f_{33} = -{[12]^2 \over (p^\perp)^4},
%\\
%& f_{12} = {[13][23] \over (p^\perp)^4}, ~~~ f_{23} = {[21][31] \over (p^\perp)^4},~~~f_{31} = {[32][12] \over (p^\perp)^4}.
%\fe
To see this, we make use of the following identity for $SU(2)$ spinors,
\ie{}
[23][11\rangle - [13][21\rangle + [12][31\rangle = 0.
\fe
It then follows that
\ie
%\sum_{i,j} f_{ij}[i1\rangle [j1\rangle =\sum_i f_{1i}[i 1\rangle =0.
\sum_k f_{ik} [kj\rangle = 0.
\fe
Then, the superamplitude can be simplified to
\ie\label{supa}
& \delta^4(P) \delta^8(Q) p^\perp \sum_{i,j} f_{ij} \theta_{iI}\theta_{jJ}\Omega^{IJ}
\\
&= \delta^4(P) {\delta^8(Q)\over (p^\perp)^3} \Big\{  [23]^2(\theta_1^2) +[31]^2(\theta_2^2)+[12]^2(\theta_3^2)
-[13][23] (\theta_1\theta_2) - [21][31] (\theta_2\theta_3) - [32][12] (\theta_3\theta_1) \Big\},
\fe
where $(\theta_i\theta_j)\equiv \theta_{iI}\theta_{jJ}\Omega^{IJ}$.
%Note that since we are working in the frame adapted to the supergraviton momentum, in the limit $p^\perp\to 0$, $p^\parallel$ which is purely along $X^0$ direction goes to zero as well, and the kinematics of $p_1, p_2, p_3$ reduces to that of a massless 3-point amplitude. In particular, $\delta^8(Q)$ vanishes like $p^\perp$ in this limit, and we see that there is indeed only a first order pole in $(p^\perp)^2$.
One can verify that, despite the $(p^\perp)^3$ in the denominator, this amplitude has only first order pole in $(p^\perp)^2$. For instance, consider the component proportional to $\eta^6\theta_1^4$, that corresponds to an amplitude that couples the 2-form potential $C_2$ in the bulk to one $+$ helicity gauge bosons and two $-$ helicity gauge bosons. This term in (\ref{supa}) scales like $\lambda_{1\A}\lambda_{1\B} [23]^2$ in our frame, which agrees with the amplitude constructed out of $F_+^2F_-^2$ vertex (in DBI action) and the 2-point $C_2 F_-$ vertex, sewn together by a gauge boson propagator, in our frame which is infinitely boosted along the momentum direction of the supergraviton. The covariantized form of this term in the superamplitude is proportional to
\ie
\delta^4(P) (\eta^6)_{AB} (\theta_1^4) {\epsilon^{IJKL} (\lambda_1^\A \zeta_{\A IA}) (\lambda_1^\B \zeta_{\B JB}) (\zeta_{\C KC}\zeta^\C{}_{LC}) [23]^2\over (p^\perp)^2}.
\fe

\begin{figure}[htb]
\centering
\begin{minipage}{0.32\textwidth}
\centering
\includegraphics[scale=2]{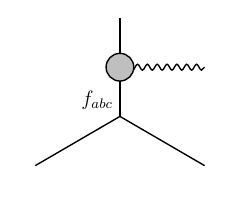}\\
\end{minipage}  
%\raisebox{-20pt}
\caption{A factorization for the $RF^3$ superamplitude for the case of an non-Abelian gauge multiplet coupled to supergravity.}
\label{nonabelian}
\end{figure}

In the case of non-Abelian gauge multiplet coupled to supergravity, there is a simpler 4-point brane-bulk superamplitude we can write down, of order $p$. The color ordered superamplitude (Figure \ref{nonabelian}) is
\ie
\delta^4(P) {\delta^8(Q) \over \langle 12\rangle \langle 23\rangle \langle 31\rangle} + ({\rm CPT ~conjugate}).
\fe
Note that this expression only has simple poles in $s_{12}$, $s_{23}$, or $s_{13}$. For instance, if we send $\langle 12\rangle\to 0$, the residue is proportional to $(p^\perp)^2$. In particular, this amplitude couples $\delta\overline\tau$ (or $\delta\tau$ from the CPT conjugate term) to three gluons of $-$ (or $+$) helicity, that factorizes through a cubic vertex in the gauge theory and a brane-bulk cubic vertex.

As another example, let us investigate a superamplitude that contains the coupling $\delta\tau F_+^2 F_-^2$. We will label the momenta of the four gauge multiplet fields $p_1,\cdots,p_4$. Such an amplitude must take the form
\ie
\delta^4(P) \delta^8(Q) {\cal F}(\lambda_{i\A},\widetilde\lambda_{i\A}),
\fe
where ${\cal F}$ is a rational function of $\lambda_i$ and $\widetilde\lambda_i$, $i=1,2,3,4$, of a total homogeneous degree $-4$ in the $\lambda_i$'s and degree 4 in the $\widetilde\lambda_i$'s. %Note that there is a component amplitude of $\delta\overline\tau$ and four $-$ helicity gauge bosons,
%\ie
%\delta^4(P) \eta^8 (p^\perp)^4 {\cal F}(\lambda_i,\widetilde\lambda_i)
%\fe 
A local supervertex would require ${\cal F}$ to be a polynomial in $\lambda,\widetilde\lambda$, which is obviously incompatible with the little group and momentum scaling. We thus conclude that there is no local supervertex that gives rise to $\delta\tau F_+^2 F_-^2$ coupling.\footnote{It appears that in string theory there is no such amplitude.}

On the D3-brane in type IIB string theory, there is a nonlocal $\delta\tau F_+^3 F_-$ amplitude. This should be part of a 5-point superamplitude of the form
\ie\label{tauF+F-3}
\delta^4(P) \delta^8(Q) \sum_{i_1,i_2,i_3,i_4}
f^{I_1I_2I_3I_4}_{i_1i_2i_3i_4}(\lambda,\widetilde\lambda) \prod_{s=1}^4 \left(\widetilde\lambda_{i_s\A}\eta^{\A}{}_{I_s} - \sqrt{p^\perp} \theta_{i_sI_s}\right) ,
\fe
where $f^{I_1I_2I_3I_4}_{i_1i_2i_3i_4}(\lambda,\widetilde\lambda)$ is a rational function of homogeneous degree $-4$ in the $\lambda$'s and degree 0 in the $\widetilde\lambda$'s.
This amplitude has a pole in $s_{123}$, $s_{124}$, $s_{134}$, $s_{234}$, and no pole in $s_{ij}$ nor in $(p^\perp)^2$. In particular, the components proportional to $\eta^8\theta_4^4$ and to $\theta_1^4\theta_2^4\theta_3^4$ (corresponding to $\delta\overline\tau F_-^3 F_+$ and $\delta\tau F_+^3 F_-$ respectively) should have only a pole in $s_{123}$. %CONSTRUCT THE SUPERAMPLITUDE EXPLICITLY.

\subsection{Soft Limits and D3-brane Coupling}

So far our considerations of brane-bulk coupling are based on supersymmetry Ward identities and unitarity of scattering amplitudes. In the context of D-branes in string theory, a crucial extra piece of ingredient is the identification of the Abelian gauge multiplet on the brane as the Nambu-Goldstone bosons and fermions associated with the spontaneous breaking of super-Poincar\'e symmetry. The amplitudes then obey a soft theorem on the scalar fields of the gauge multiplet. The soft theorem relates the amplitude ${\cal A}(\phi^{IJ},\cdots)$ with the emission of a Nambu-Goldstone boson $\phi^{IJ}$ in the soft limit to the amplitude ${\cal A}(\cdots)$ without the $\phi^{IJ}$ emission,
\ie
\lim_{p_{\phi}\to 0}{\cal A}(\phi^{IJ},\cdots) = \sqrt{g\over T} \, p^{IJ} {\cal A}(\cdots).
\fe
Here $p^{IJ}$ is the $[IJ]$-component of the total momentum transverse to the 3-brane. The normalization of the soft factor is unambiguously determined by the relation between ${\cal B}_{1,1}$ and the 1-point amplitude ${\cal B}_1$.

Let us consider the 3-point amplitude between a supergraviton and two gauge multiplets. The momenta of the two gauge multiplets and the graviton are $p_1, p_2, p_3$, with $p_1+p_2+p_3^\parallel=0$.
The amplitude takes the form
\ie
{\cal B}_{1,2} = g {\delta^8(Q)\over \langle12\rangle^2}.
\label{B12}
\fe
%Note that the amplitude ${\cal B}_{1,2}$ is in fact local, because $\delta^8(Q)$ scales with the momentum like $t^2$, ${\cal B}_{1,2}$ scales like $t$, and there is no pole in $t$. 
Expanding in components, we have
\ie
{\cal B}_{1,2} = g\Big( [12]^2 \eta_3^8 + \langle 12\rangle^2 \theta_1^4\theta_2^4 + \cdots \Big),
\fe
where the terms proportional to $\theta_1^4\theta_2^4$ and $\eta_3^8$  give the vertices for $\tau F_+^2$ and $\bar\tau F_-^2$ coupling, respectively. Note that $(p_3^\perp)^2 = -(p_1+p_2)^2 = -2p_1\cdot p_2 = \langle 12\rangle [12]$.

\begin{figure}[htb]
\centering
%\begin{minipage}{0.49\textwidth}
%\centering
\includegraphics[scale=2 ]{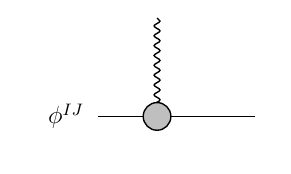}
%\end{minipage}  
\raisebox{45pt}{$\xrightarrow{p_\phi\rightarrow 0}$}
%\begin{minipage}{0.5\textwidth}
%\centering
\includegraphics[scale=2]{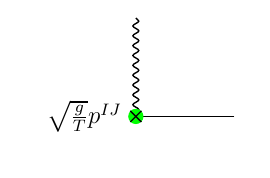}
%\end{minipage} 
\caption{Single soft limit of ${\cal B}_{1,2}$. }
\label{soft}
\end{figure}

${\cal B}_{1,2}$ is related to ${\cal B}_{1,1}$ by taking the soft limit on a scalar $\phi_{IJ}$ on the brane (Figure \ref{soft}). The soft theorem on the Nambu-Goldstone bosons $\phi_{IJ}$ implies that, in the limit $p_1\to 0$,
\ie
\left.{\cal B}_{1,2}\right|_{\theta_{1I}\theta_{1J}} \to \sqrt{g\over T}\, p^{IJ} {\cal B}_{1,1},
\fe
where $p^{IJ}=p_3^{IJ}$ is the $\phi^{IJ}$ component of the transverse momentum. More explicitly, we can write
\ie\label{bsoft}
\left.{\cal B}_{1,2}\right|_{\theta_{1I}\theta_{1J}} = g {\lambda_{1\A}\lambda_{1\B}\over \langle 12\rangle^2 } \, \delta^{6(\A\B)[IJ]}(q_2+q_3),
\fe
where
\ie
\delta^{6(\A\B)[IJ]}(Q) &= {1\over 768}\Big[\epsilon^{II_1I_2I_3}(Q^\A{}_{I_1} Q_{\A_1I_2}Q_{\A_2I_3}) \epsilon^{JJ_1J_2J_3}  (Q^\B{}_{J_1} Q^{\A_1}{}_{J_2}Q^{\A_2}{}_{J_3}) 
\\
&~~~~+ \epsilon^{I_1I_2I_3I_4}(Q^\A{}_{I_1} Q^\B{}_{I_2} Q^{\A_1}{}_{I_3} Q^{\A_2}{}_{I_4}) \epsilon^{IJ J_1J_2} (Q_{\A_1J_1} Q_{\A_2J_2})\Big].
\fe
The RHS of (\ref{bsoft}), after imposing $p_2+p_3^\parallel=0$, is independent of the choice of $\lambda_1$, and is proportional to the 2-point bulk-brane vertex ${\cal B}_{1,1}$.

More specifically, let us choose the frame as in the supervertex $\mathcal{B}_{1,1}$. We take $p_2,\, p_3^\parallel$ to be along a lightlike direction in the $(X^0,X^1)$ plane and $p_3^\perp$ to be along a lightlike direction on the $(X^8,X^9)$ plane. The  $SO(6)$ spinor indices $I$ is broken into spinor indices $a,\,\dot a$ of $SO(4)$ that rotates $X^4,\,X^5,\,X^6,\,X^7$. We pick the transverse momentum of the supergraviton to be along the direction $[IJ]=[ab]$ on the $(X^8,X^9)$ plane (while $[IJ]=[a\dot b]$ would be a direction in the $X^4,\,X^5,\,X^6,\,X^7$ space). The spinor helicity variables in this frame are given by \eqref{frame}. In particular, $\lambda_{2+}=\sqrt{p^\parallel_+}, \lambda_{2-}=0$ and $p_3^{IJ} = p^\perp _+$. Focusing on the $(\alpha,\beta)=(-,-)$ term in \eqref{bsoft}, this is indeed proportional to the supervertex $\mathcal{B}_{1,1}$ in the soft limit in this frame:
\ie
g {\lambda_{1+}\lambda_{1+} \over \langle 12\rangle^2}\delta^{6(--)[ab]}(q_2+q_3)
\propto g {\delta^{6}(Q_{+a}, Q_{-a}, Q_{+\dot a})\over p^{\parallel }_+} = \sqrt{g\over T}\,p^\perp_+\mathcal{B}_{1,1}.
\fe

\subsection{The Brane-Bulk Effective Action}

Let us comment on the notion of effective action for the brane in our consideration of higher derivative couplings. We will be interested in the ``massless open string 1PI" effective action for a D3-brane in type IIB string theory. Namely, we will be considering a quantum effective action through which the full massless open-closed string scattering amplitudes are reproduced by sewing effective vertices through ``disc type" tree diagrams, that is, diagrams that correspond to factorization through either massless open or closed string channels of a disc diagram. 

This effective action is subject to two subtleties. The first is the appearance of non-analytic terms. This is familiar in the massless closed string effective action already: in type IIB string theory, there are for instance string 1-loop non-analytic terms at $\A' D^2 R^4$ and $\A'^4 D^8 R^4$ order in the momentum expansion. Often, the higher derivative terms one wishes to constrain does not receive non-analytic contributions in the quantum effective action of string theory. Sometimes, when the non-analytic terms do appear, such as those of the same order in momentum as $D^2 R F^2$ and $R^2$ terms in the D3-brane effective action, as will be discussed in the next section, their effect is to add a term that is linear in the dilaton (logarithmic in $\tau_2$) to the coefficient of the higher derivative coupling of interest, which is related to a modular anomaly.

If we work with a Wilsonian effective action, take the floating cutoff $\Lambda$ to be very small (compared to string scale) and then consider the momentum expansion, the non-analytic term is absent, and instead of the $\log\tau_2$ contribution, we will have a constant shift of the coefficient of the higher derivative operator (like $D^2 RF^2$ or $R^2$) that depends logarithmically on $\Lambda$. Our analysis of supersymmetry constraints applies straightforwardly in this case (and as we will see, such constant shifts are compatible with supersymmetry). In doing so, however, one loses the exact $SL(2,\mathbb{Z})$ invariance in the effective coupling, and the modular anomaly must be taken into account to recover the $SL(2,\mathbb{Z})$ symmetry.

\begin{figure}[htb]
\centering
\begin{minipage}{0.49\textwidth}
\centering
\includegraphics[scale=1.8 ]{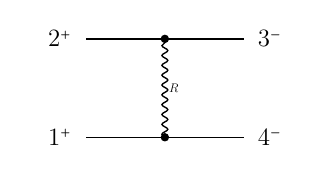}
\end{minipage}  
\begin{minipage}{0.49\textwidth}
\centering
\includegraphics[scale=1.8 ]{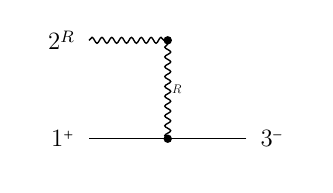}
\end{minipage} 
\caption{Examples of non-disc type diagrams. The black dots represent (bare) brane-bulk coupling. }
\label{loop}
\end{figure}

The second subtlety has to do with the brane. Note that, in the ``massless open string 1PI" effective action, closed string propagators that connect say a pair of discs have been integrated out already. This is because the tree diagrams that involves bulk fields connecting pairs of brane vertices behave like loop diagrams (Figure \ref{loop}), where the transverse momentum of the bulk propagator is integrated \cite{Goldberger:2001tn,Michel:2014lva}. Therefore, in analyzing tree level unitarity of superamplitudes built out of higher derivative vertices of the effective action, we will consider only the ``disc type" tree diagrams.

\section{Supersymmetry Constraints on Higher Derivative Brane-Bulk Couplings}

Following a similar set of arguments as in \cite{Wang:2015jna}, we will derive non-renormalization theorems on $f_F(\tau,\bar\tau) F^4$ terms that couple the Abelian field strength on the brane to the dilaton-axion of the bulk type IIB supergravity multiplet, and on $f_{RFF}(\tau,\bar\tau) RF^2$ and $f_R(\tau,\bar\tau) R^2$ terms that couple the brane to the bulk dilaton-axion and graviton.

\subsection{$F^4$ Coupling}

Let us suppose that there is supersymmetric $F^4$ coupling on the brane, whose coefficient $f_F(\tau,\bar\tau)$ depends on the axion-dilaton field $\tau$ in the bulk. Consider a vacuum in which the dilaton-axion field $\tau$ acquires expectation value $\tau_0$, and we denote its fluctuation by $\delta\tau$. Expanding
\ie\label{expff}
f_F(\tau,\bar\tau) F^4 = f_F(\tau_0,\bar\tau_0) F^4 + \partial_\tau f_F(\tau_0,\bar\tau_0)\delta\tau F^4
+ \partial_{\overline\tau} f_F(\tau_0,\bar\tau_0)\delta\overline\tau F^4 + \partial_\tau \partial_{\overline\tau} f_F(\tau_0,\bar\tau_0)\delta\tau\delta\overline\tau F^4 + \cdots,
\fe
one could ask if the coefficient of $\delta \tau F^4$, namely $\partial_\tau f_F$ at $\tau=\tau_0$, is constrained by supersymmetry in terms of lower point vertices. This amounts to asking whether the coupling $\delta\tau F^4$ admits a local supersymmetric completion, as a supervertex. As already argued in the previous section, such a supervertex does not exist. The reason is that the desire supervertex, in $\theta$-representation, must be of the form
\ie
\delta^4(P)\delta^8(Q_\theta) {\cal F}(\lambda_i,\bar\lambda_i),
\fe
where ${\cal F}(\lambda_i,\widetilde\lambda_i)$ must have total degree $-4$ in $\lambda_i$, $i=1,\cdots,4$, and degree 4 in $\widetilde\lambda_i$, as constrained by the little group scaling on the massless 1-particle states in four dimensions. Such a rational function will necessarily introduce poles in the Mandelstam variables, and will not serve as a local supervertex.

The situation is in contrast with the 4-point $F^4$ supervertex, which does exist. There, the rational function ${\cal F}$ can be written as $[34]^2/\langle 12\rangle^2$, which due to the special kinematics of 4-point massless amplitude in four dimensions does not introduce poles in momenta. This is not the case for higher than 4-point amplitudes, where the local supervertex of the similar form does not exist. Also note that, had there been such a 5-point supervertex, it would give rise to an independent $\delta\tau F_+^2F_-^2$ coupling, whereas in string theory the analogous nonlocal superamplitude on the D3-brane contains an amplitude of the form $\delta\tau F_+^3 F_-$ instead.

\begin{figure}[htb]
\centering
\includegraphics[scale=2]{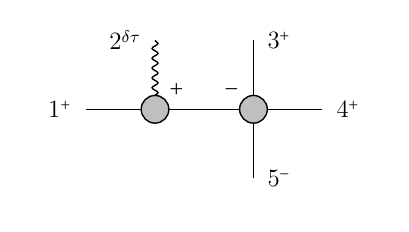}
\caption{Factorization of the $\D\tau F_+^3 F_-$ amplitude through one $F_+^2F_-^2$ vertex and an $RF^2$ supervertex.}
\label{tauF+3F-}
\end{figure}

Now that an independent $\delta\tau F^4$ supervertex does not exist, the coefficient $\partial_\tau f_F$, which is given by the soft limit of a 5-point superamplitude, is fixed by the residues of the 5-point superamplitude at its poles. It must then be fixed by lower point supervertices, namely, by the coefficient of $F^4$. This means that there is a linear relation between $\partial_\tau f_F$ and $f_F$, which takes the form of a first order differential equation on $f_F(\tau,\bar\tau)$. In fact, as noted already below \eqref{tauF+F-3}, the actual 5-point superamplitude that factorizes through an $F^4$ supervertex has degree 12 in $\eta$ and $\theta$ (see Figure \ref{tauF+3F-}), so the $\delta\tau F_+^2 F_-^2$ coupling which has degree 8 in $\eta$ and $\theta$ must not be part of this superamplitude and the first order differential equation simply says that $f_F(\tau,\bar\tau)$ is a constant. 

This is indeed what we see in the DBI action for a D3-brane in type IIB string theory. In the usual convention, the gauge kinetic term is normalized as $\tau_2 F^2$, and the DBI action contains $\tau_2 F^4$ coupling in string frame, which translates into $\tau_2^2 F^4$ in Einstein frame \cite{Johnson:2000ch}. In the consideration of scattering amplitudes, it is natural to rescale the gauge field by $\tau_2^{-1/2}$, so that the kinetic term is canonically normalized. This is the correct normalization convention in which the expansion (\ref{expff}) applies, and the DBI action corresponds to $f_F(\tau,\bar\tau)=1$. Thus, we conclude that the tree level $F^4$ coupling is exact in the full quantum effective action of type IIB string theory. Note that, rather trivially, this result is consistent with $SL(2,\mathbb{Z})$ invariance. Unlike the $R^4$ coupling in type IIB string theory, however, here the constraint from supersymmetry is stronger, and one need not invoke $SL(2,\mathbb{Z})$ to fix the $F^4$ coefficient.

The above discussion is in contrast to the $F^4$ coupling in the Coulomb phase of a four dimensional gauge theory with sixteen supersymmetries.\footnote{We restrict our discussion to the rank 1 case. The spacetime dimension of the gauge theory is not essential here.} In this case, one may consider the $F^4$ coefficient as a function of the scalar fields on the Coulomb branch moduli space. There are independent supervertices of the form 
\ie
\delta^4(P)\delta^8(\mathcal{Q}_\psi)
\fe
in the $\psi$-representation, that contains couplings of the form $\phi^{i_1}\cdots \phi^{i_n} F^4+\cdots$ and transforms in the rank $n$ symmetric traceless tensor representation of the $SO(6)$ R-symmetry. As a consequence, through consideration of factorization of 6-point superamplitudes at a generic point on the Coulomb branch, one derives a second order differential equation that asserts $\Delta_\phi f(\phi)$ is proportional to $f(\phi)$. Comparison with DBI action then fixes this differential equation to simply the condition that $f(\phi)$ is a harmonic function. This reproduces the result of \cite{Paban:1998ea,Paban:1998qy}.

\subsection{$RF^2$ Coupling}

The 3-point superamplitude between one supergraviton and two gauge multiplets is particularly simple because there is only one invariant Mandelstam variable, $t=(p_3^\perp)^2 = \langle 12\rangle [12]$, where $p_3$ is the momentum of the supergraviton. A general 3-point superamplitude of this type takes the form (in $\theta$-representation)
\ie\label{asd}
{\cal A}_{1,2} = \delta^4(P) {\delta^8(Q_\theta)\over \langle 12\rangle^2} f(t),~~~f(t) = \sum_{n\geq -1} f_n t^{n+1}.
\fe
Previously, we have considered the term $f_{-1}$ which we called ${\cal B}_{1,2}$ in \eqref{B12}. We have seen that it is not renormalized, and is fixed by the bulk cubic coupling. We will work in units in which this coupling is set to 1. Now let us consider the possibility of having $f_n$ for general $n\geq 0$ as a function of the dilaton-axion $\tau,\bar\tau$. 

First, let us ask what are the independent local supervertices that could couple $\delta\tau, \delta\overline\tau$ to $RF^2$. Such an $(3+m)$-point supervertex, with the correct little group scaling in four dimensions, must take the form
\ie
\delta^4(P) \delta^8(Q_\theta) {{\cal P}_{n+1} \over \langle 12\rangle^2},
\fe
where ${\cal P}_{n+1}$ is a function of the spinor helicity variables that scales with momentum like $t^{n+1}$. For $m\geq 1$, the $\langle 12\rangle^2$ in the denominator must be canceled by a factor from the numerator in order for the supervertex to be local (there is no longer the special kinematic constraint as in the case of the 3-point vertex that renders (\ref{asd}) local even for the $f_{-1}$ term). For this, we need $n\geq 1$, so that we can write a local supervertex of the form
\ie\label{fvert}
\delta^4(P)\delta^8(Q_\theta) [12]^2 {\cal P}_{n-1}.
\fe
The 4-point superamplitude for $\tau R F_+ F_-$ can not factorize through lower point supervertices. It follows that the coefficient $f_0$ in (\ref{asd}) as a function of $\tau,\bar\tau$ is subject to a homogenous first order differential equation, which simply states that $f_0$ is a constant. Moreover as we shall see below, $f_0$ is fixed to be identically zero using tree-amplitude in type IIB string theory.

\begin{figure}[htb]
\centering
\begin{minipage}{0.49\textwidth}
\centering
\includegraphics[scale=1.7 ]{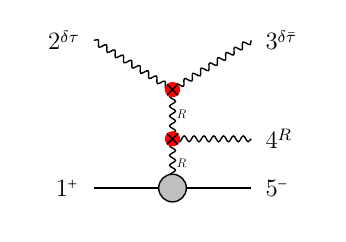}
\end{minipage}  
\begin{minipage}{0.49\textwidth}
\centering
\includegraphics[scale=1.7]{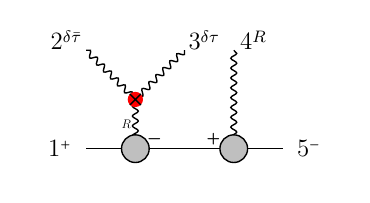}
\end{minipage} 
\caption{Factorization of the $\D\tau \D\bar{\tau} R F_+ F_-$ amplitude through lower-point vertices.}
\label{tautaubarRFF}
\end{figure}

Supervertices of the form (\ref{fvert}) are F-term vertices, and give rise to $(\delta\tau)^m D^{2n} RF^2$ coupling.
We would like to constrain $\partial_\tau \partial_{\bar\tau} f_n$ from supersymmetry, by showing that as the coefficient of a coupling of the form $\delta\tau\delta\overline\tau D^{2n} RF^2$, it cannot be adjustable by introducing a local supervertex. So let us focus on the 5-point supervertices.
When $n\geq 2$, such a coupling may be part of a 5-point D-term supervertex of the form
\ie\label{5ptDtermRF2}
\delta^8(Q) \overline Q^8 {\cal F}(\lambda_i,\widetilde\lambda_i,\theta_i,\zeta_j,\eta_j),
\fe
where ${\cal F}$ is of homogeneous degree $2(n-2)$ in the momenta. For $n=1$, on the other hand, the only available supervertex is the F-term vertex of the form (\ref{fvert}), which gives $(\delta\tau)^2D^2 RF^2$ rather than $\delta\tau\delta\overline\tau D^2 RF^2$ coupling. There appears to be no independent 5-point supervertex for $\delta\tau\delta\overline\tau D^2 RF^2$, and the supersymmetric completion of such a coupling can only be a nonlocal superamplitude. Therefore, $f_1$ is determined by the factorization of the 5-point superamplitude into lower point superamplitudes, that involves 1 or 2 cubic vertices of the type $f_0$ or $f_1$ (Figure \ref{tautaubarRFF}). Thus, we have relations of the form
\ie\label{tre}
4\tau_2^2\partial_\tau \partial_{\bar\tau} f_1(\tau,\bar\tau) = a f_1 + b f_0^2,
\fe
where $a,b$ are constants that are fixed entirely by tree level unitarity and supersymmetry Ward identities.

Let us compare this with the disc amplitude on D3-branes in type IIB string theory, where $f(t)$ is given by (in string frame) \cite{Hashimoto:1996bf}
\ie
-2{\Gamma(-2t)\over \Gamma(1-t)^2} = t^{-1} + \zeta(2) t+ 2\zeta(3) t^2 + \cdots,
\fe
which, after going to Einstein frame and rescaling the gauge field so that the gauge kinetic term is canonically normalized, corresponds to
\ie\label{stringtreeRF2}
f_{-1} = 1,~~~ f_0 = 0,~~~ f_1 = \zeta(2) \tau_2,~~~ f_2 = 2\zeta(3) \tau_2^{3/2}, ~~~etc.
\fe
As remarked earlier, $f_0=0$ is an exact result in the full quantum effective action for the D3-brane in type IIB string theory. Comparing with (\ref{tre}), we learn that $f_1(\tau,\bar\tau)$ is a harmonic function on the axion-dilaton target space. Knowing its asymptotics in the large $\tau_2$ limit, we can then determine this function by $SL(2,\mathbb{Z})$ invariance.

There is a subtlety here, having to do with non-analytic terms from the open string 1-loop amplitude, that gives rise to a $\log(\tau_2) D^2 RF^2$ term. As a consequence, $f_1(\tau,\bar\tau)$ is only $SL(2,\mathbb{Z})$ invariant up to an additive modular anomaly. This is similar to the modular anomaly of the $R^2$ coefficient, pointed out in \cite{Bachas:1999um,Basu:2008gt} and to be discussed below. 
After taking into account the modular anomaly, $f_1$ is unambiguously fixed to be%\footnote{} 
\ie
f_1(\tau,\bar\tau) = {1\over 2} Z_1(\tau,\bar\tau) = \zeta(2) \tau_2 -{\pi\over 2} \ln\tau_2 + \pi \sum_{m,n=1}^\infty {1\over n} \left( e^{2\pi i mn\tau}+e^{-2\pi i mn\bar\tau}\right).
\fe
Here we denote the non-holomorphic Eisenstein series by $Z_s=2\zeta(2s)E_s$ \cite{Green:1999pv},
\ie
Z_s=\sum_{(m,n)\neq (0,0)} {\tau_2^s\over |m+n\tau|^{2s}},
\fe
which have the weak coupling expansion (for $s\neq 1$),
\ie
Z_s=&2\zeta(2s)\tau_2^s+2\sqrt{\pi}\tau_2^{1-s}{\Gamma(s-1/2)\zeta(2s-1)\over \Gamma(s)}+{\cal O}(e^{-2\pi \tau_2}).
%\\&+{2\pi^s\over \Gamma(s)}\sum_{k\neq 0}\m(k,s)
%e^{-2\pi (|k|\tau_2-ik\tau_1)}|k|^{s-1}\left(1+{s(s-1)\over 4\pi |k| \tau_2}+\dots\right)
\fe

For $n=2$, the candidate 5-point D-term supervertex \eqref{5ptDtermRF2} has an ${\cal F}$ which is of degree $0$ in the momenta. In order to achieve the correct little group scaling for $D^4RF^2$, ${\cal F}$ must be a non-constant function of $[12]/\la12\ra$  which would lead to a nonlocal expression in the absence of special kinematics. Therefore we conclude there's no independent $\D\tau  \D\bar\tau D^4RF^2$ supervertex, which again results in a 2nd order differential equation of the form,
\ie
4\tau_2^2\partial_\tau \partial_{\bar\tau} f_2(\tau,\bar\tau) = a f_2 (\tau,\bar\tau)
\fe
where we've used $f_0=0$. String tree level amplitude \eqref{stringtreeRF2} fixes $a=3/4$. Combining with $SL(2,\bZ)$ invariance, we have $f_2=E_{3/2}$. In particular, the perturbative contributions to $D^4 RF^2$ come from only open string tree-level and two-loop orders.

\subsection{$R^2$ Coupling on the Brane}

\begin{figure}[htb]
\centering
\begin{minipage}{0.33\textwidth}
\centering
\includegraphics[scale=1.5]{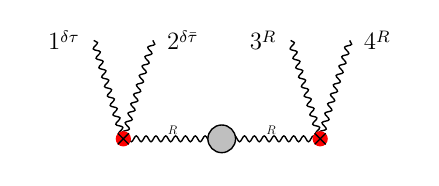}
\end{minipage} \hfill
\begin{minipage}{0.33\textwidth}
\centering
\includegraphics[scale=1.5]{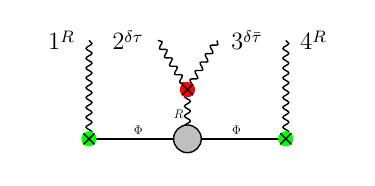}
\end{minipage}\hfill
\begin{minipage}{0.33\textwidth}
\centering
\includegraphics[scale=1.5]{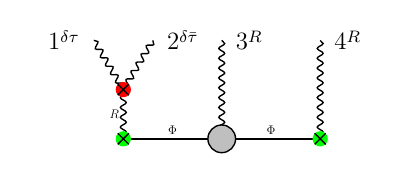}
\end{minipage}
\caption{Potential factorizations of the $\D\tau \D\bar{\tau} R^2$ amplitude through lower-point vertices.}
\label{tautaubarRR}
\end{figure}

Now we turn to $R^2$ coupling on the 3-brane. The F-term supervertices for $n$-point super-graviton coupling to the brane at four-derivative order are given by
\ie
\delta^4(P_\m)\delta^8(\sum_{i=1}^n Q_{i \A I})= \delta^4(P_\m)\delta^8\left( \sum_{i=1}^n \xi_{i\A I}{}^A \eta_{iA} \right)
\fe
and its CPT conjugate. Since there are no four-dimensional particles involved in this amplitude, there is no little group scaling to worry about. These F-term vertices contain $\delta\tau^{n-2} R^2$ and $\delta\overline\tau^{n-2} R^2$ couplings. The mixed $\delta\tau^n\delta\overline\tau^m D^{2k} R^2$ couplings, as part of a local supervertex, can come from D-term supervertices for $k\geq 2$, but not for $k=0,1$. The $\delta\tau\delta\overline\tau R^2$ coupling can only be the soft limit of a 4-point brane-bulk superamplitude, that factorizes through either an $R^2$ vertex or a $D^2 RF^2$ vertex, along with the elementary vertices (Figure \ref{tautaubarRR}).\footnote{A priori, the 4-point brane-bulk superamplitude could factorize through two $RF^2$ type vertices, giving rise to a source term in the differential constraint proportional to $f_0^2$. However as argued before, $f_0=0$ holds to all orders.} The coefficient of $\delta\tau\delta\overline\tau R^2$ is determined by the residues at these poles, thereby related linearly to $R^2$ and $D^2RF^2$ coefficients. We immediately learn that the coefficient $f_R(\tau,\bar\tau)$ of $R^2$ coupling must obey
%\footnote{It is insufficient to rule out certain factorization channels by focusing on a particular component of the superamplitude.}
\ie\label{rteq}
4\tau_2^2 \partial_\tau \partial_{\overline\tau} f_R(\tau,\bar\tau) = a f_R(\tau,\bar\tau) + b f_1(\tau,\bar\tau),
\fe
where $f_1(\tau,\bar\tau)$ is the coefficient of $D^2 RF^2$.

%Moreover, focusing on the $\D\tau^2\D\bar{\tau}^2$ component of the same 4-point superamplitude, we see that the absence of coupling to dilaton in the 2-point supervertex ${\cal B}_{1,1}$ eliminates the factorization channels through an $RF^2$ supervertice (last two diagrams in Figure \ref{tautaubarRR}). This forces $b$ to vanish.  Now if $a$ is nonzero, comparison with the string tree level answer \cite{Hashimoto:1996bf} then implies that $f_R$ cannot have an order $\tau_2$ term, and its perturbative expansion in $\tau_2^{-1}$ only contains non-positive powers of $\tau_2$. On the other hand, writing $a = s(s-1)$, then the eigen-modular function $f_R$ must have perturbative terms of order $\tau_2^s$ and $\tau_2^{1-s}$, which would lead to a contradiction unless this function is identically zero.  In conclusion, $f_R(\tau,\bar\tau)$ is a harmonic function, and since it should be a modular function modulo the modular anomaly due to a $\log\tau_2$ term coming from the non-analytic terms in the quantum effective action, it is given by the modular completion of its asymptotic expansion at large $\tau_2$, namely $Z_1(\tau,\bar\tau)$. This proves the conjecture of \cite{Basu:2008gt}.

Let us compare this relation with the perturbative results in type IIB string theory. In the previous subsection we have fixed $f_1(\tau,\bar\tau)$ to be ${1\over 2} Z_1(\tau,\bar\tau)$. $f_R$ receives the contribution $2\zeta(3) \tau_2$ from the disc amplitude \cite{Hashimoto:1996bf}. This gives a linear relation between $a$ and $b$. Modulo the modular anomaly due to non-analytic terms, $f_1$ is a harmonic function, and so $a f_R+bf_1$ is either zero (which implies that $f_R$ is harmonic) or an eigenfunction of the Laplacian operator with eigenvalue $a$. If $a$ is zero, the equation (\ref{rteq}) is incompatible with the tree level result of $f_1$. If $a$ is nonzero, comparison with the tree level answer then implies that $af_R + bf_1$ cannot have an order $\tau_2$ term, and its perturbative expansion in $\tau_2^{-1}$ only contains non-positive powers of $\tau_2$. On the other hand, writing $a = s(s-1)$, then the eigen-modular function $af_R+bf_1$ must have perturbative terms of order $\tau_2^s$ and $\tau_2^{1-s}$, which would lead to a contradiction unless this function is identically zero.  In conclusion, $f_R(\tau,\bar\tau)$ is also a harmonic function, and since it should be a modular function modulo the modular anomaly due to a $\log\tau_2$ term coming from the non-analytic terms in the quantum effective action, it is given by the modular completion of its asymptotic expansion at large $\tau_2$, namely $Z_1(\tau,\bar\tau)$. This proves the conjecture of \cite{Basu:2008gt}.

\begin{figure}[htb]
\centering
\begin{minipage}{0.49\textwidth}
\centering
\includegraphics[scale=1.7]{ttbRR2.pdf}
\end{minipage}  
\begin{minipage}{0.49\textwidth}
\centering
\includegraphics[scale=1.7 ]{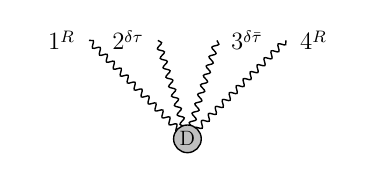}
\end{minipage} 
\caption{A factorization channel of the $\delta\tau \bar\delta \tau D^4 R^2$ amplitude and a $D$-term supervertex that contributes at the same order.}
\label{D4R2}
\end{figure}

In a similar way, we can derive the supersymmetry constraint on $D^2 R^2$ coupling. The independent $D^2 R^2$ supervertices are
\ie
\delta^4(P) \delta^8(Q_{1aI}+Q_{2aI}) s_{12}^\perp,\quad \delta^4(P) \delta^8(Q_{1aI}+Q_{2aI}) u_{12},
\fe
where $s_{12}^\perp = - (p_1^\perp + p_2^\perp)^2$ and $u_{12}=-4(p_1^\perp)^2+(p_1^\perp + p_2^\perp)^2$,
%u_{12}=2(p_1^\perp\cdot p_2^\perp-(p_1^\perp)^2)$
$p_i^\perp$ being the component of the momentum of the $i$-th particle perpendicular to the 3-brane. F-term $n$-point supervertices give rise to $\delta\tau^{n-2} D^2 R^2$ and $\delta\overline\tau^{n-2} D^2 R^2$ couplings, but $\delta\tau\delta\overline\tau D^2 R^2$ coupling is not part of a local supervertex, and must be the soft limit of a 4-point superamplitude that factorizes through the $D^2 R^2$ vertex. Note that the first D-term supervertex that contributes to the 4-point amplitude starts at the order of $D^4 R^2$ (Figure \ref{D4R2}), and would not affect the $D^2 R^2$ superamplitude. Thus the independent coefficients $f^s_{R,2}(\tau,\bar\tau)$  and $f^u_{R,2}(\tau,\bar\tau)$ of $D^2 R^2$ supervertex obey a second order differential equation of the form
\ie\label{ffd}
4\tau_2^2 \partial_\tau \partial_{\overline\tau} \begin{pmatrix}
f^s_{R,2}(\tau,\bar\tau) \\ f^u_{R,2}(\tau,\bar\tau)
\end{pmatrix}  = M \begin{pmatrix}
f^s_{R,2}(\tau,\bar\tau) \\ f^u_{R,2}(\tau,\bar\tau)
\end{pmatrix},~~M\in {\rm Mat}_{2\times 2}(\bR) .
\fe
By comparing with the $D^2 R^2$ term in the disc and annulus 2-graviton amplitude on a D3-brane in type IIB string theory, which is proportional to $\tau_2^{3/2}u_{12} R^2 (1+{\cal O}(\tau^{-2}))$  in Einstein frame\footnote{
%The absence of open string one-loop contribution to the superampltitude of order $D^2R^2$ can be understood as follows. 
The open string annulus diagram involves gauge multiplets in the loop joined by two (bare) brane-bulk supervertices of the type $RF^2$. However the absence of $RF^2$ supervertex at order $p^4$ implies that the open string annulus diagram gives no contribution to the two point superamplitude of order $D^2 R^2$.} \cite{Hashimoto:1996bf,Basu:2008gt}, we conclude that $M$ has an eigenvector  $\left(\begin{smallmatrix} 0 \\1 \end{smallmatrix}\right)$ with eigenvalue $3/4$. Combined with $SL(2,Z)$-invariance, this allows us to determine $f^u_{R,2}=Z_{3/2}$ up to an nonzero constant coefficient. Now the other independent differential constraint is $4\tau_2^2\partial_{\tau}\partial_{\bar \tau} f^s_{R,2}=a f^s_{R,2}+b f_{R,2}^u$. If $b\neq 0$, the leading contribution to $f^s_{R,2}$ in $\tau_2^{-1}$ must be $\tau_2^{3/2}\log\tau_2$ up to a nonzero constant, but such non-analytic piece cannot appear at tree level in string perturbation theory. Writing $a=s(s-1)$, then $f^s_{R,2}$ is an eigen-modular function with perturbative terms of order $\tau_2^s$ and $\tau_2^{1-s}$. However since $f^s_{R,2}$ receives no contribution at order $\tau^{3/2}$ (tree) and $\tau^{1/2}$ (open string one loop), consistency of string perturbation theory demands $f^s_{R,2}=0$ identically. To sum up, the $D^2R^2$ coupling on the brane is captured by a single eigen-modular function $f^u_{R,2}=Z_{3/2}(\tau,\bar{\tau})$.

\section{Torus Compactification of 6D $(0,2)$ SCFT}

Let us consider the six dimensional $A_{N-1}$ $(0,2)$ superconformal theory compactified on a torus of modulus $\tau$, to a four dimensional quantum field theory that may be viewed as the $SU(N)$ ${\cal N}=4$ super-Yang-Mills theory, deformed by higher dimensional operators that preserve 16 supercharges and $SO(5)\subset SO(6)$ R-symmetry. We would like to determine these higher dimensional operators.

\subsection{Harmonicity Condition on the Coulomb Branch Effective Action}

A clear way to address this question is to consider the Coulomb phase of the theory, and study the effective action of Abelian gauge multiplets. We will focus on couplings of the form
\ie
f(\tau,\overline\tau, \phi_i, y) F^4,
\fe
where $\phi_i$, $i=1,\cdots,5$ and $y$ constitute the six scalars $\Phi_i$ in the gauge multiplet, with the $\phi_i$ transforming in the vector representation of $SO(5)$. We may view the compactification as first identifying the 6D $A_1$ $(0,2)$ SCFT compactified on circle with a 5D gauge theory, which is 5D maximally supersymmetric $SU(2)$ gauge theory up to D-term deformations, and then further compactifying the 5D gauge theory \cite{Seiberg:1997ax,Douglas:2010iu,Lambert:2010iw}. On the Coulomb branch, the scalar $y$ comes from the Wilson line of the Abelian gauge field, and is circle valued. 

It is known from \cite{Paban:1998ea} that the $(\phi_i, y)$ dependence is such that $f(\tau,\overline\tau, \phi_i, y)$ is a harmonic function on the moduli space $\mathbb{R}^5\times S^1$. In the amplitude language, as already explained in section 2, this can be argued as follows. Expanding near a point on the Coulomb branch, the only supervertices of the form $(\delta\phi)^2 F^4$ are in the symmetric traceless representations of the local $SO(6)$ R-symmetry, whereas the R-symmetry singlet $(\delta\phi)^2 F^4$ coupling can only be part of a nonlocal amplitude. Unlike the supergravity case, here the Coulomb branch effective theory would be free without the $F^4$ and higher derivative couplings, and the six point amplitude can only factorize into a pair of $F^4$ or higher order supervertices, and in particular cannot have polar terms at the same order in momenta as $(\delta\phi)^2 F^4$. It follows that the $SO(6)$ singlet $(\delta\phi)^2 F^4$ vertex is absent, which is equivalent to the statement that $f(\tau,\bar\tau,\phi_i, y)$ is annihilated by the Laplacian operator on the Coulomb moduli space.
The $(\tau,\overline\tau)$ dependence of the $F^4$ coupling, on the other hand, does not follow from supersymmetry constraints on the low energy effective theory. 

As a side comment, if we start with M-theory on a torus that is a product of two circles of radii $R_{10}$ and $R_9$, wrap M5-branes on the torus times $\mathbb{R}^{1,3}$, reduce to type IIA string theory along the circle of radius $R_{10}$ and T-dualize along the other circle, we obtain D3-branes in type IIB string theory with $\tau=i R_{10}/R_9$, compactified on a circle of radius 
\ie
\widetilde R = \ell_s {\ell_{11}^{3\over 2}\over R_9 R_{10}^{1\over 2}} 
\fe
that is transverse to the D3-branes. Here $\ell_{11}$ is the 11 dimensional Planck length and $\ell_s$ is the string length. To identify the four dimensional world volume theory with the torus compactification of the $(0,2)$ SCFT requires taking the limit $R_9,R_{10} \gg \ell_{11}$, which implies that $\widetilde R\ll \ell_s$. Thus, it is unclear whether the four dimensional gauge theory of question can be coupled to type IIB supergravity, with $\tau$ identified with the dilaton-axion field.

\subsection{Interpolation through the Little String Theory}

Nonetheless, without consideration of coupling to supergravity, we will be able to determine the function $f(\tau,\overline\tau,\phi_i, y)$ completely (including the $\tau,\overline\tau$ dependence) by an interpolation in the Coulomb phase of the torus compactified $(0,2)$ little string theory, in a similar spirit as in \cite{Lin:2015zea}. 
Based on the $SO(5)$ symmetry and the harmonicity of $f(\tau,\overline\tau, \phi_i, y)$, we can put it in the form
\ie\label{crho}
f(\tau,\overline\tau, \phi_i, y) = c(\tau,\bar\tau) + \sum_{n\in\mathbb{Z}} \int_0^{2\pi{\cal R}} dv {\rho(\tau,\overline\tau, v)\over \left[ |\phi|^2 + (y-v -2\pi n {\cal R} )\right]^2}.
\fe
Here $2\pi {\cal R}$ is the periodicity of the field $y$. The constant term $c(\tau,\overline\tau)$ and the source profile $\rho(\tau,\overline\tau, v)$ are yet to be determined functions. Now let us compare this to the Coulomb branch effective action of the $A_1$ $(0,2)$ little string theory (LST) compactified on a torus, of complex modulus $\tau$ and area $L^2$. The Coulomb moduli space ${\cal M}_{LST}$ is parameterized by the expectation values of four scalars $\phi_i$, $i=1,\cdots,4$, a fifth compact scalar $\phi_5$, and the zero mode of the self-dual 2-form potential $A = {1\over 2} A_{\mu\nu}dx^\mu\wedge dx^\nu$, namely
\ie
y = L^{-1} \int_{T^2} A.
\fe
Here we defined $y$ such that it has a canonically normalized kinetic term, and has periodicty $L^{-1}$($\equiv 2\pi{\cal R}$). The compact scalar $\phi_5$, on the other hand, has periodicity $L/\ell_s^2$.\footnote{This comes from the zero mode of a six dimensional compact scalar of periodicity $1/\ell_s^2$, normalized with canonical kinetic term in four dimensions.} 
The torus compactified $(0,2)$ superconformal theory is obtained in the limit $\ell_s\to 0$ while keeping $L$ finite. In this limit $\phi_5$ decompactifies while $y$ retains the periodicity $L^{-1}$.

Far away from the origin on the Coulomb branch, the $(0,2)$ LST can be described by the double scaled little string theory, whose string coupling $g_s$ is related to the expectation values of the scalar fields $\phi_i$ (after compactification to four dimensions) through\footnote{To see this identification, we go back to NS5-branes in type IIA string theory, separated in the transverse $\mathbb{R}^4$ by the displacement $\vec x$. The double scaled little string theory (DSLST) is defined by the limit $|\vec x|\to 0$, holding $g_{eff} = g_s^\infty\ell_s/|x|$ fixed, where $g_s^\infty$ is the asymptotic string coupling before taking the decoupling limit. $g_{eff}$ is then identified with the string coupling at the tip of the cigar in the holographic description of DSLST, which we denote by $g_s$. After further compactifying the DSLST on a torus of area $L^2$ to four dimensions, our normalization convention on the scalar fields $\phi_i$ and $y$ is such that $y$ is identified with $(g_s^\infty \ell_s L)^{-1}$ times the displacement of the 5-branes along the M-theory circle, while $\phi_i$ is identified with $(g_s^\infty \ell_s L)^{-1}x_i$. This then fixes the normalization in the relation between $g_s$ of DSLST and $|\phi|$.}
\ie
g_s = {1\over L\sqrt{ \sum_{i=1}^4 \phi_i^2}}.
\fe
%In the weak coupling limit $g_s\rightarrow0$, and therefore large $|\phi_i|$, the $F^4$ term in the Coulomb effective action can be computed from DSLST perturbation theory, and must fall off like $1/|\phi|^2$ at large $\phi_i$ without a constant term.
 Together with the $SO(4)$ symmetry and harmonicity condition on $\mathbb{R}^4\times T^2$, the coefficient of $F^4$ in LST should take the form 
\ie\label{flst}
&f_{LST}(\tau,\overline\tau, \phi_i, y)= %{1\over \pi \ell_s^2} 
 c(\tau,\overline\tau , L/\ell_s)+
\sum_{n,m\in\mathbb{Z}} \int du dv {\rho(\tau,\overline\tau, L/\ell_s, u, v) \over \left[ \sum_{i=1}^4 \phi_i^2 + (\phi_5 - u -  m L/\ell_s^2)^2 + (y - v - n/L)^2  \right]^2},
\fe
where $u, v$ are integrated along the $\phi_5$ and $y$ circles in the moduli space. In the weak coupling limit $g_s\rightarrow0$, and therefore large $|\phi_i|$ with  $i=1,\cdots,4$, the $F^4$ term in the Coulomb effective action can be computed reliably from the LST perturbation theory. In particular,  in the large $\phi_i$ limit, the leading contribution to $f_{LST}$ comes from the tree level scattering amplitude, which scales like $g_s^2 \sim |\phi|^{-2}$, plus corrections of order $e^{-|\phi|}$.\footnote{Note that, from the DSLST perspective, there are no higher order perturbative contributions to $f_{LST}$, but there are non-perturbative contributions. It would be interesting to recover these non-perturbative terms of order $e^{-|\phi|}$ by a D-instanton computation in the $(0,2)$ DSLST on the torus.} This then fixes the constant term $c(\tau,\overline\tau,L/\ell_s)$ to be zero and 
\ie
\int du dv \,\rho(\tau,\overline\tau,L/\ell_s, u,v) = 1,
\fe
which is in particular \textit{independent} of $\tau,\overline\tau$.

%Note that, the integral of the source $\rho$ is determined by a tree level gauge boson scattering amplitude in double scaled LST, which is independent of the modulus $\tau$ of the compactification torus. In the large $\phi_i$ limit ($i=1,\cdots,4$), $f_{LST}$ is proportional to $|\phi|^{-2}$, plus corrections of order $e^{-|\phi|}$.\footnote{Note that, from the DSLST perspective, there are no higher order perturbative contributions to $f_{LST}$, but there are non-perturbative contributions. It would be interesting to recover these non-perturbative terms of order $e^{-|\phi|}$ by a D-instanton computation in the $(0,2)$ DSLST on the torus.} Matching the coefficient of $|\phi|^{-2}$ with tree level double scaled LST fixes $\int du dv \,\rho(\tau,\overline\tau,L/\ell_s, u,v) = 1$, which is in particular independent of $\tau,\overline\tau$.

In the limit 
\ie\label{02limit}
\ell_s\to 0,~~~~L, ~\phi_1,\cdots,\phi_5,~y~{\rm finite},
\fe
the $(0,2)$ LST reduces to the $(0,2)$ superconformal theory, and we should recover $SO(5)$ R-symmetry. In this limit, the $F^4$ coefficient \eqref{flst} becomes a harmonic function on $\mathbb{R}^5\times S^1$,  thus the source $\rho$ in (\ref{flst}) should be localized at $u=0$. This argument also determines $c(\tau,\overline\tau)=0$ in (\ref{crho}). Next, if we further take the limit 
\ie
L\to 0,~~~~\phi_1,\cdots,\phi_5,~y~{\rm finite},
\fe
we should recover four dimensional ${\cal N}=4$ SYM, where the higher dimensional operators (to be discussed below) are suppressed, with the $SO(6)$ R-symmetry restored. 
%The underlying conformal invariance of the $(0,2)$ SCFT implies that the distribution of the source $\rho$ along the $y$-circle is independent of $L$,
In this limit, the coefficient $f(\tau,\overline\tau,\phi_i,y )$ for the $F^4$ term becomes a harmonic function on $\mathbb{R}^6$, 
  so we learn that $\rho$ must be supported at $v=0$ as well. Importantly, as stated below (\ref{flst}), the matching with tree level DSLST amplitudes at large $|\phi|$ fixes the overall normalization of $\rho$ to be independent of $\tau,\bar\tau$, hence $\rho(\tau,\bar\tau,\infty, u,v) = \delta(u)\delta(v)$.
Thus, we determine $f(\tau,\overline\tau,\phi_i, y)$ to be given {\it exactly} by (after rescaling all scalar fields by $L/(2\pi)$)
\ie\label{hfunc}
H(\phi_i, y) = \sum_{n\in\mathbb{Z}} {1\over \left[ |\phi|^2 + (y-2\pi n)^2 \right]^2}
\fe
as the coefficient of $F^4$ in the Coulomb branch of the $A_1$ $(0,2)$ SCFT. \footnote{The periodicity of $y$ is either $4\pi$ or $2\pi$ depending on whether the gauge group is $SU(2)$ or $SO(3)$. Here we are considering the case of $SO(3)$ where there is a single singularity on the moduli space where the $SO(6)$ R-symmetry is restored. The $SU(2)$ case will be considered in \cite{xiyin}.}

The key to the above argument is that while the dependence on $\tau$, which are the complexified coupling constant, of the torus compactified $(0,2)$ theory could \textit{a priori} be arbitrarily complicated, the dependence on $\tau$, which becomes the modulus of the target space torus, of the LST tree level scattering amplitude is completely trivial. By interpolating between the weakly coupled $(0,2)$ LST with the $(0,2)$ superconformal field theory, we determine the $\tau$ dependence of the $F^4$ coefficient of the latter theory.

We have implicitly worked in the convention where the gauge fields have canonically normalized kinetic terms. If we work in the more standard field theory convention where the kinetic term for the gauge field is written as $\tau_2 F^2$, then the $F^4$ term acquires a factor $\tau_2^2$, and so we can write
\ie\label{ones}
f(\tau,\overline\tau,\phi_i, y) = \tau_2^2 H(\phi_i, y).
\fe

Let us compare this with our expectation in the large $\tau_2$ regime, where $F^4$ coupling can be computed from 5D maximal SYM compactified on a circle, by integrating out $W$-bosons that carry Kaluza-Klein momenta at 1-loop. As argued in \cite{Lin:2015zea}, the 5D gauge theory obtained by compactifying the $(0,2)$ SCFT (as opposed to little string theory) does not have ${\rm tr} F^4$ operator at the origin of the Coulomb moduli space, thus the 1-loop result from 5D SYM holds in the large $\tau_2$ regime. This indeed reproduces (\ref{ones}).

Near the origin of the Coulomb branch, expanding in $\phi_i$ and in $y$, the term $n=0$ in (\ref{hfunc}) can be understood as the 1-loop $F^4$ term in the Coulomb effective action of ${\cal N}=4$ SYM. The $n\not=0$ terms, which are analytic in the moduli fields at the origin, can be viewed as $F^4$ and higher dimensional operators that deform the ${\cal N}=4$ SYM at the origin. From the expansion
\ie
\sum_{n\not=0} {1\over \left[ |\phi|^2 + (2\pi n-y)^2 \right]^2}
= {\zeta(4)\over 8\pi^4} + {\zeta(6)\over 16\pi^6} \left(5 y^2-|\phi|^2\right) + {\zeta(8)\over 128\pi^8} \left[ 35 y^4 - 42 y^2|\phi|^2 + 3(|\phi|^2)^2\right] + \cdots
\fe
we can read off the operators at the origin of the moduli space,\footnote{Our result (\ref{opex}) disagrees with the proposal of \cite{Douglas:2010iu}, where a different modular weight was assigned to $f(\tau,\bar\tau)$, and the proposed answer has a subleading perturbative term in $\tau_2^{-1}$. One can directly verify, from the circle compactification of 5D SYM, that there are no higher loop contribution to the $F^4$ term through integrating out KK modes. The higher loop corrections to the effective action only appear at $D^2 F^4$ order and above. Furthermore, by unitarity cut construction it appears that the $F^4$ term in the Coulomb effective action cannot be contaminated by higher dimensional operators in the 5D gauge theory that come from the compactification of $(0,2)$ SCFT.}
\ie\label{opex}
{\zeta(4)\over 8\pi^4} \tau_2^2 {\cal O}^{(8)} + { 3 \zeta(6)\over 8\pi^6} \tau_2^2 {\cal O}^{(10)}_{66} + \cdots
\fe
Here ${\cal O}^{(8)}$ is the $1/2$ BPS dimension 8 operator that is the supersymmetric completion of ${\rm tr} F^4$, whereas ${\cal O}^{(10)}_{ij}$ is the $1/2$ BPS dimension 10 operator in the symmetric traceless representation of $SO(6)$ R-symmetry, of the form
\ie
{\cal O}^{(10)}_{ij} = {\rm tr} (\Phi_{(i} \Phi_{j)} F^4) - {1\over 6} \delta_{ij} {\rm tr} (|\Phi|^2 F^4) + \cdots
\fe
Likewise, there is a series of higher dimensional $1/2$ BPS operators that transform in higher rank symmetric traceless representations of the R-symmetry. In fact, these are all the BPS (F-term) operators that are Lorentz invariant in the $SU(2)$ maximally supersymmetric gauge theory. In the higher rank case, i.e. torus compactification of $A_r$ $(0,2)$ SCFT for $r>1$, the 4D gauge theory is also deformed by the $1/4$ BPS dimension 10 double trace operator of the form $D^2 {\rm tr}^2 F^4+\cdots$, and analogous higher dimensional operators in nontrivial representations of R-symmetry. These receive contributions from the circle compactified 5D SYM at two-loop order.

\bigskip

\section*{Acknowledgments}

We would like to thank Clay C\'ordova and Thomas Dumitrescu for discussions. Y.W. is supported in part by the U.S. Department of Energy under grant Contract Number  DE-SC00012567. X.Y. is supported by a Sloan Fellowship and a Simons Investigator Award from the Simons Foundation.

\bibliography{nonrenormrefs} 
\bibliographystyle{JHEP}
\end{document}